\documentclass[twocolumn,trackchanges]{aastex7}
\usepackage[T1]{fontenc}
\usepackage{stackengine}

\shorttitle{Multiwavelength Analysis of Winged Radio Galaxies}
\shortauthors{Bera et al.}
\accepted{for publication in ApJ}
\begin{document}

\title{A Morphological Identification and Study of Radio Galaxies from LoTSS DR2. III. The multiwavelength analysis of Winged Radio Galaxies}

\correspondingauthor{Taotao Fang}
\email{fangt@xmu.edu.cn}

\author[orcid=0000-0001-7121-4258,gname='Soumen Kumar',sname='Bera']{Soumen Kumar Bera}
\affiliation{Department of Astronomy, Xiamen University, Xiamen, Fujian 361005, People's Republic of China}
\email{beras@xmu.edu.cn}
\author[orcid=0000-0002-2853-3808,gname='Taotao',sname='Fang']{Taotao Fang}
\affiliation{Department of Astronomy, Xiamen University, Xiamen, Fujian 361005, People's Republic of China}
\email{fangt@xmu.edu.cn}
\author[0000-0002-6741-9856,gname='Magdalena',sname='Kunert-Bajraszewska']{Magdalena Kunert-Bajraszewska}
\affiliation{Institute of Astronomy, Faculty of Physics, Astronomy and Informatics, NCU, Grudziądzka 5/7, 87-100, Toruń, Poland}
\email{magda@astro.umk.pl}
\author[orcid=0000-0001-9251-9456,gname='Tapan K.',sname='Sasmal']{Tapan K. Sasmal}
\affiliation{National Astronomical Observatories, Chinese Academy of Sciences, Beijing 100101, People's Republic of China}
\email{tapan.phys@gmail.com}
\author[orcid=0000-0002-3462-4175,gname='Si-Yue',sname='Yu']{Si-Yue Yu}
\affiliation{Kavli Institute for the Physics and Mathematics of the Universe (Kavli IPMU, WPI), UTIAS, Tokyo Institutes for Advanced Study, University of Tokyo, Chiba 277-8583, Japan}
\affiliation{Department of Astronomy, Xiamen University, Xiamen, Fujian 361005, People's Republic of China}
\email{si-yue.yu@ipmu.jp}
\author[orcid=0000-0003-1761-5442,gname='Chuan',sname='He']{Chuan He}
\affiliation{Department of Astronomy, Xiamen University, Xiamen, Fujian 361005, People's Republic of China}
\email{hechuan@xmu.edu.cn}
\author[orcid=0000-0001-6475-8863,gname='Xuelei',sname='Chen']{Xuelei Chen}
\affiliation{National Astronomical Observatories, Chinese Academy of Sciences, Beijing 100101, People's Republic of China}
\email{xuelei@bao.ac.cn}
\author[gname='Soumen',sname='Mondal']{Soumen Mondal}
\affiliation{Department of Physics, Jadavpur University, Kolkata 700032, India}
\email{crabhorizon@gmail.com}

\begin{abstract} 
We present a multiwavelength follow-up study of 621 winged radio galaxies (WRGs) recently identified from LoTSS DR2, constituting the largest statistically significant samples of X-shaped (XRGs) and Z-shaped (ZRGs) radio galaxies to date. Our results show that WRGs are predominantly strongly radio-dominated, with XRGs on average more radio-luminous than ZRGs. Their optical hosts are massive elliptical galaxies residing in moderate-density environments.
For 270 of XRGs, we measure angular offsets between the radio wings and the optical major axis. While most XRGs show large misalignments consistent with hydrodynamic backflow along the host minor axis, a substantial fraction (~25\%) exhibits small offsets (<30°), indicating that additional processes, such as jet reorientation, may also play a role. ZRGs, in contrast, are characterized by strongly antisymmetric deformations of their radio lobes pointing toward a coherent mechanism affecting both jets, modulated by local environmental interactions at the lobe termini.
Mid-infrared diagnostics indicate merger-related cold gas in many WRGs, particularly XRGs, which also more frequently host powerful AGN, while ZRGs are more often classified as low-excitation radio galaxies (LERGs). This is consistent with our previous results showing that, although most WRGs exhibit FR II morphologies, FR I sources are almost exclusively ZRGs, suggesting that Z-shaped structures are statistically associated with lower jet power and are therefore more susceptible to perturbations.
Nevertheless, the physical processes responsible for shaping XRGs and ZRGs need not be fundamentally different. Instead, the final morphology likely reflects the interplay between jet power, jet stability, and the surrounding environment.

\end{abstract}

\section{Introduction}
\label{sec:intro}
Winged radio galaxies (WRGs) represent a fascinating class of irregular radio galaxies, exhibiting an additional pair of diffuse, low-surface-brightness lobes called as wings, along with the collimated, high-surface-brightness primary lobes \citep{1984MNRAS.210..929L, 1992ersf.meet..307L, 2025ApJS..278...34B}. The WRGs account for a fraction of 3\% to 10\% of radio galaxies \citep{1992ersf.meet..307L, 2019ApJ...887..266J} and largely are Fanaroff-Riley class II \citep[FR II;][]{1974MNRAS.167P..31F} sources \citep{1984MNRAS.210..929L, 2025ApJS..278...34B}. This class of sources is further subdivided into two classes, namely, X-shaped radio galaxies \citep[XRGs;][]{2007AJ....133.2097C, 2019ApJS..245...17Y} and Z-shaped radio galaxies \citep[ZRGs;][]{2005MNRAS.364..583Z}, depending on the ejection point of the wings and overall morphology \citep{2020ApJS..251....9B, 2025ApJS..278...34B}. The morphological difference between these two sub-groups might reflect different physical mechanisms or environmental interactions \citep{2003ApJ...594L.103G}.

The underlying mechanism of the genesis of winged radio sources is still a subject of active debate. In general, all of the models are divided between ambient medium influences and intrinsic central engine dynamics \citep{2010ApJ...720L.155G, 2024FrASS..1171101G}. In the extrinsic, environmental models, the hydrodynamical backflow model is a popular model where the synchrotron plasma is assumed to flow backward due to the asymmetric circumgalactic medium, creating the wings \citep{1984MNRAS.210..929L, 2002A&A...394...39C, 2011ApJ...733...58H, 2017A&A...606A..57R}. Among the intrinsic models, it is considered that events like sudden jet reorientation or spin-flip are possibly due to supermassive black hole (SMBH) mergers \citep{2001A&A...377...23Z, 2002Sci...297.1310M} or accretion disk instabilities \citep{2002MNRAS.330..609D}. There is also another consideration of the presence of dual AGN at the core \citep{2005MNRAS.356..232L, 2007MNRAS.374.1085L}. In spite of these models and the respective studies, there is no single formation model that is capable of explaining the entire population of WRGs \citep{2019ApJ...887..266J, 2022ApJS..260....7B, 2024FrASS..1171101G}.

Until recently, the majority of the identifications of winged sources were limited to the VLA-FIRST survey at 1.4 GHz (e.g., \citet{2007AJ....133.2097C, 2019ApJS..245...17Y, 2020ApJS..251....9B}). Hence, the further studies are also somewhat biased towards the relatively high frequency, bright, or classical X-shaped sources (e.g., \citet{2010MNRAS.408.1103L, 2011ApJ...733...58H, 2018ApJ...852...47R}). So, in our previous work \citet{2025ApJS..278...34B} (hereafter \hyperlink{cite.2025ApJS..278...34B}{Paper I}) we used the second data release of LOFAR Two-metre Sky Survey  \citep[LoTSS DR2;][]{2022A&A...659A...1S} at 144 MHz to conduct an extensive search for WRGs. By using the combination of high sensitivity ($83~\rm {\mu}Jy~beam^{-1}$) and excellent resolution ($6^{\prime\prime}$) of LoTSS DR2, we made the largest identification of winged sources to date, with 621 bona fide, along with an additional 403 candidates. The 621 WRGs include 382 XRGs and 239 ZRGs. A few basic parameters, like two-point spectral index ($\alpha_{144}^{1400}$), radio power at 144 MHz ($P_{144 MHz}$), and FR classification, were also studied.

While the morphological identification of the WRGs is an important first step, the understanding of these sub-class of sources and their underlying origin requires a detailed investigation of their intrinsic properties at different wavelengths. There are a few recent studies that attempted to address this topic using a moderate-sized sample (e.g., \citet{2015ApJS..220....7R, 2016A&A...587A..25G, 2019ApJ...887..266J, 2020ApJS..251....9B}, and all are from VLA-FIRST data). These works primarily focus on the radio luminosity, spectral index, optical host properties, and local environment. However, regarding the formation models, the works \citep{2002A&A...394...39C, 2009ApJ...695..156S, 2016A&A...587A..25G, 2019ApJ...887..266J} particularly focus on the radio and optical structural alignment. This is due to the fact that the prior evidence suggests that the wings in XRGs have a tendency to align with the optical minor axis of their elliptical hosts, which supports the hydrodynamical backflow model \citep{2011ApJ...733...58H, 2022A&A...662A...5G}. In the studies, the X-shape feature is primarily considered.

The large sample of WRGs identified in LoTSS DR2 offers a unique opportunity for a comprehensive multiwavelength investigation. Since XRGs and ZRGs may arise from distinct formation mechanisms or environmental conditions, we examine their properties separately whenever possible. Building on \hyperlink{cite.2025ApJS..278...34B}{Paper I}, this work presents a systematic multiwavelength study of the WRG population.

The paper is structured as follows. Section~\ref{sec:sample} describes the WRG sample. Section~\ref{sec:radio_prop} presents the radio properties, including position angle and radio loudness. Section~\ref{sec:optical_prop} details the optical host parameters, such as ellipticity, host position angle, radio–optical misalignment, stellar mass, environmental density metrics, and cluster associations. Section~\ref{sec:infrared_prop} examines the infrared hosts, while Section~\ref{sec:xray_prop} summarizes the available X-ray detections. A discussion of the results and their implications for wing formation models is given in Section~\ref{sec:summary}. Throughout this work we adopt a cosmology with $\rm H_0 = 67.4$ km s$^{-1}$ Mpc$^{-1}$, $\rm \Omega_m = 0.315$, and $\rm \Omega_{\Lambda} = 0.685$ \citep{2020A&A...641A...6P}.

\begin{figure*}[ht!]
\centering
\centering
\stackinset{l}{21pt}{t}{5pt}{\includegraphics[width=2.30cm]{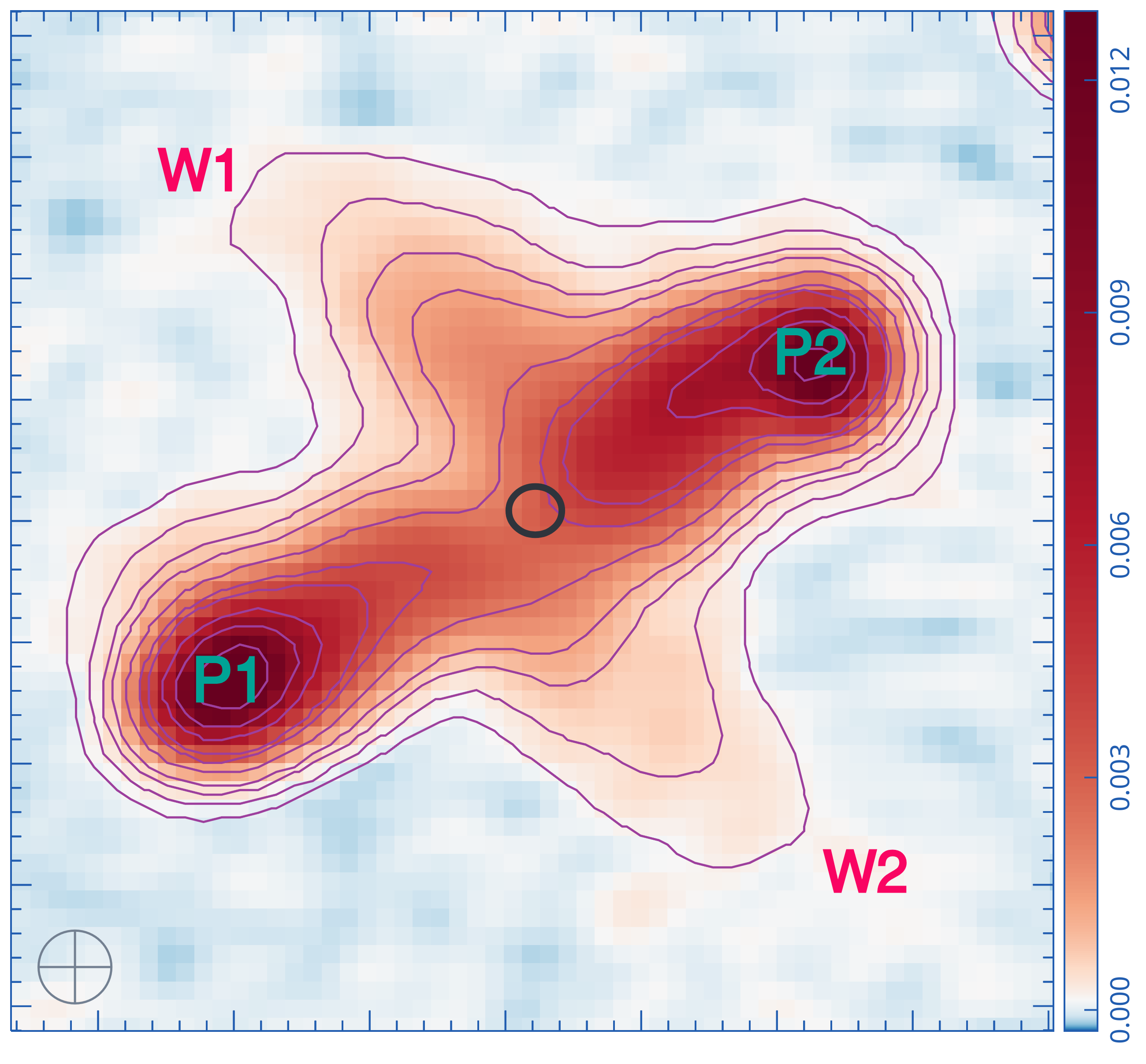}}{\includegraphics[width=11.50cm]{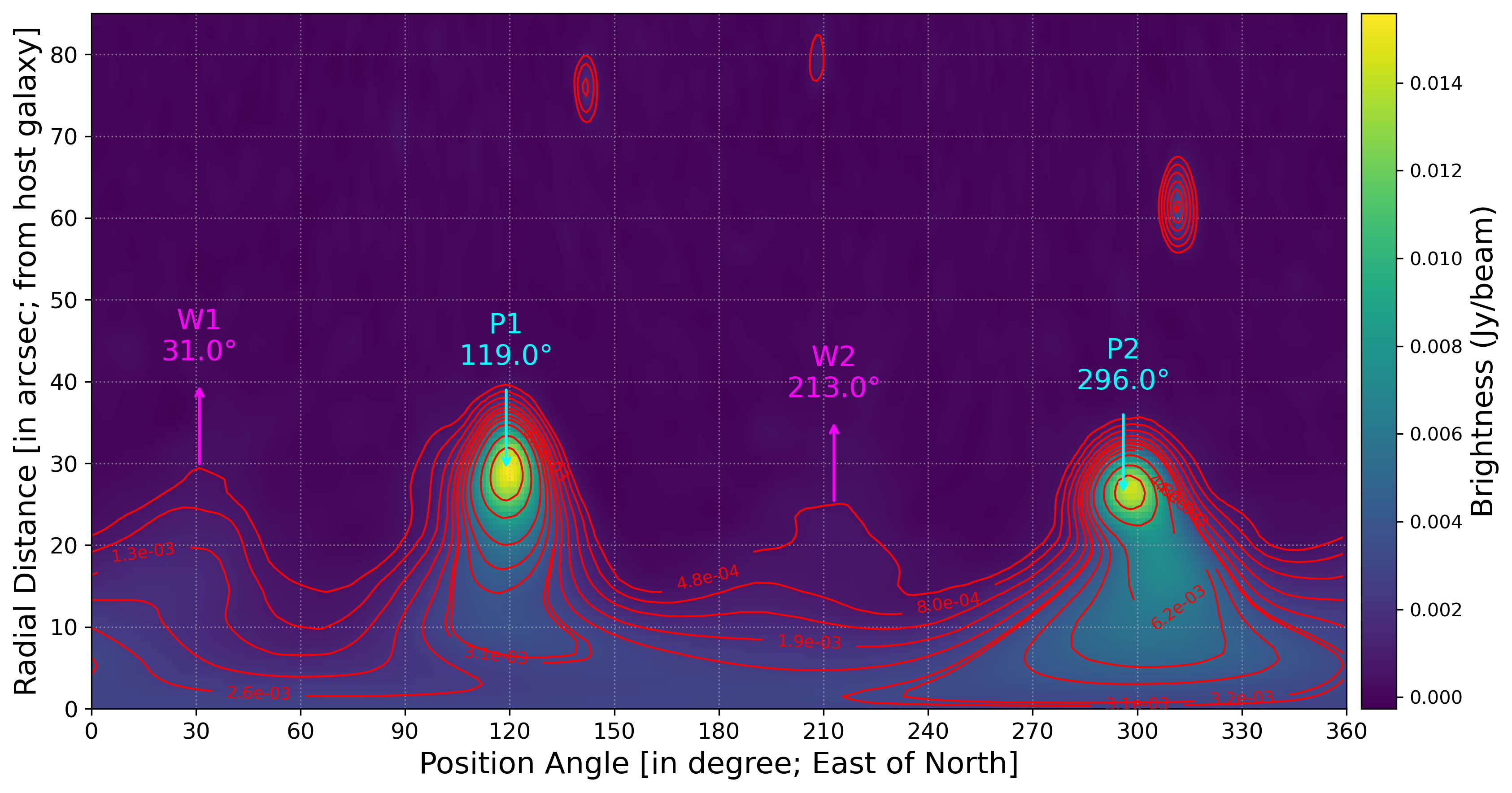}}
\includegraphics[width=6.40cm]{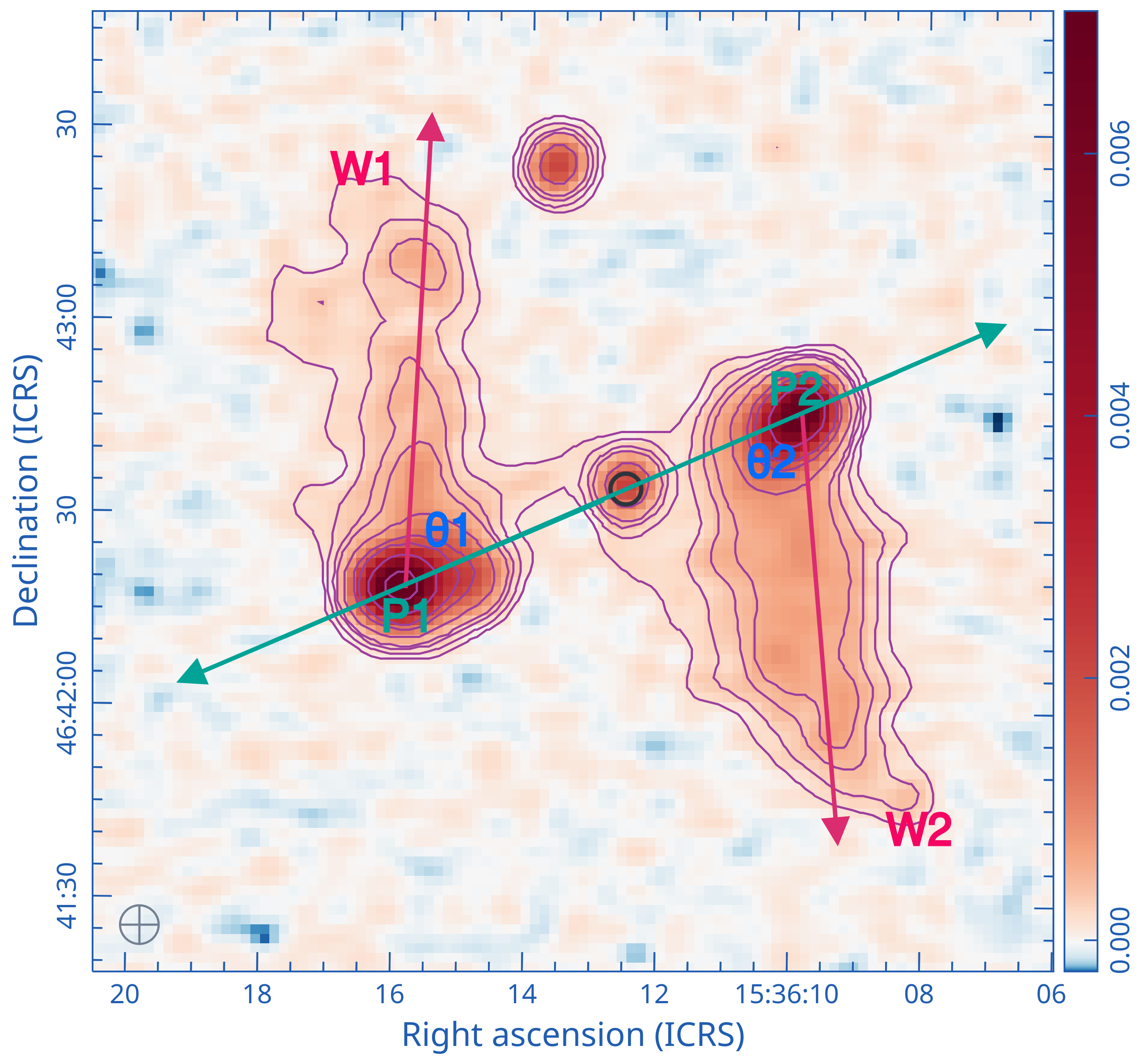}
\label{fig:radio_PA}
\caption{Examples of radio position angle measurements for XRGs and ZRGs.
\textit{Left:} Measurement of the primary lobe and wing PAs for an XRG using the polar map method. The inset shows the LoTSS DR2 radio image of XRG J0959+4146, with the primary lobes (P1, P2), wings (W1, W2), and the host galaxy marked by a central circle. The main panel displays the corresponding polar map constructed by taking the host galaxy as the origin. Inward-pointing cyan arrows indicate the PAs of the primary lobes, while outward-pointing magenta arrows mark the PAs of the wings.
\textit{Right:} Schematic illustrating the measurement of wing–lobe offset angles for a ZRG. The angles $\theta1$ and $\theta2$ are defined between the axes of the primary lobes (P1, P2) and their corresponding wings (W1, W2), as shown for the ZRG J1536+4642.}
\end{figure*}

\section{The Sample}
\label{sec:sample}
With reference to our earlier work (\hyperlink{cite.2025ApJS..278...34B}{Paper I}), the present study investigates the multiwavelength properties of WRGs. In that study, we reported a total of 1024 winged sources, including 403 candidates. Here, we restrict our analysis to the bona fide sample of 621 confirmed winged sources, which serves as the starting point of this investigation.

A prerequisite for the multiwavelength analysis is the secure identification of the host galaxies. In \hyperlink{cite.2025ApJS..278...34B}{Paper I}, the optical hosts were adopted from \citet{2023A&A...678A.151H}, who used the 16th Data Release of the Sloan Digital Sky Survey \citep[SDSS DR16;][]{2020ApJS..249....3A}. However, given the extensive size of the SDSS catalog ($\sim$4 million sources), occasional misidentifications of proposed optical counterparts cannot be ruled out. To minimize this risk, we re-examined each WRG and independently verified its host galaxy association within SDSS. As a result, we confirmed reliable optical counterparts for 453 sources—281 XRGs and 172 ZRGs—from the 521 optical hosts identified previously in \hyperlink{cite.2025ApJS..278...34B}{Paper I}. This re-identification was performed using the more recent SDSS DR18 \citep{2023ApJS..267...44A}.

It is important to emphasize that not all sources have complete or reliable coverage across every wavelength. Objects lacking adequate data quality or clear detections in a given band were excluded from the corresponding analyses, leading to variations in sample size depending on the specific wavelength regime. These adjustments are described explicitly within the relevant sections.

\section{Radio Properties}
\label{sec:radio_prop}

\subsection{Radio Position Angle}
\label{subsec:radio_pa}
Owing to the intrinsic morphological differences between XRGs and ZRGs, we adopt different position angle (PA) measurement schemes for the wings in the two classes, tailored to their physical interpretation. Figure~\ref{fig:radio_PA} illustrates these two approaches using representative examples of an XRG and a ZRG. The resulting radio PAs are subsequently used in the position angle offset analysis presented in Section~\ref{sec:pa_offset}. All PAs are measured east of north.

For XRGs, where the wings emerge from the central region of the host galaxy, the PAs of both the primary radio lobes and the wings are measured with respect to the host position. This is achieved using polar maps constructed from the LoTSS DR2 radio images. For each source, the host galaxy is taken as the origin and the radio image is transformed into polar coordinates following the method described in \citet{2016A&A...587A..25G}. An illustrative example is shown in the left panel of Figure~\ref{fig:radio_PA}. In the resulting polar diagrams, the PA of each primary lobe is defined by the angle corresponding to the peak radio brightness, while the PA of each wing is determined as the direction of maximum radial extent at which the emission remains above the local $3\sigma$ noise level (as in \hyperlink{cite.2025ApJS..278...34B}{Paper I}). This procedure robustly recovers the orientations of both components, as demonstrated for the XRG J0959+4146.

In contrast, for ZRGs the wings do not originate near the host galaxy but instead emerge from the outer edges of the primary lobes. According to our classification scheme, ZRGs are defined as WRGs in which the secondary lobes emerge beyond one-fourth of the primary lobe length (see \hyperlink{cite.2025ApJS..278...34B}{Paper I}, Section 3.3). Equivalently, by defining the `Z-length' as the projected distance along the primary axis between the two wing emergence points, ZRGs satisfy the condition of Z-length/primary lobe size $>$ 0.25. This operational definition ensures a clear separation from XRGs. Consequently, for this class we characterize the wing orientation relative to the axis defined by the primary radio lobes, rather than with respect to the optical host. Specifically, the offset angle is measured between each primary lobe (P1, P2) and its corresponding wing (W1, W2). An example of this measurement scheme is shown in the right panel of Figure~\ref{fig:radio_PA} for the ZRG J1536+4642.

In both classes, the two primary lobes (and similarly the two wings) are not always perfectly antiparallel. We therefore define representative PAs for the lobes and wings by averaging the angles of the individual components. While this averaging can introduce small uncertainties—particularly in morphologically complex sources or those with very diffuse wings (examples are presented in Appendix~\ref{appendix_PA_measurements})—its impact on the global statistics is limited, and it remains the most practical approach for characterizing orientation offsets in a large sample. Adopting a tolerance of $10^\circ$ to classify lobe pairs as aligned, we find that approximately 25\% of sources exhibit significant lobe misalignment. The reliability of our PA measurements is further assessed through a comparison with literature values, as described in Appendix~\ref{appendix_PA_verification}.

\subsection{Radio Loudness}
Radio loudness (R) compares the power output of an object in radio wavelengths to its power output in optical wavelengths, and it helps to analyze the radio-loud/radio-quiet (RL/RQ) dichotomy for a source. Here we adopted the formula given by \citet{2019A&A...622A..11G} as:

\begin{equation}
R = \log_{10}\left(\frac{L_{radio}}{L_{optical}}\right) = \log_{10}\left(\frac{L_{\mathrm{144\,MHz}} / \mathrm{W\ Hz^{-1}}} {L_{\mathrm{i-\,band}} / \mathrm{W\ Hz^{-1}}}\right)
\end{equation}

where $\rm L_{144\,MHz}$ is the LoTSS 144 MHz radio luminosity and $\rm L_{i-\,band}$ is the i-band optical luminosity.

\begin{figure}[ht!]
\includegraphics[angle=0,width=8.50cm]{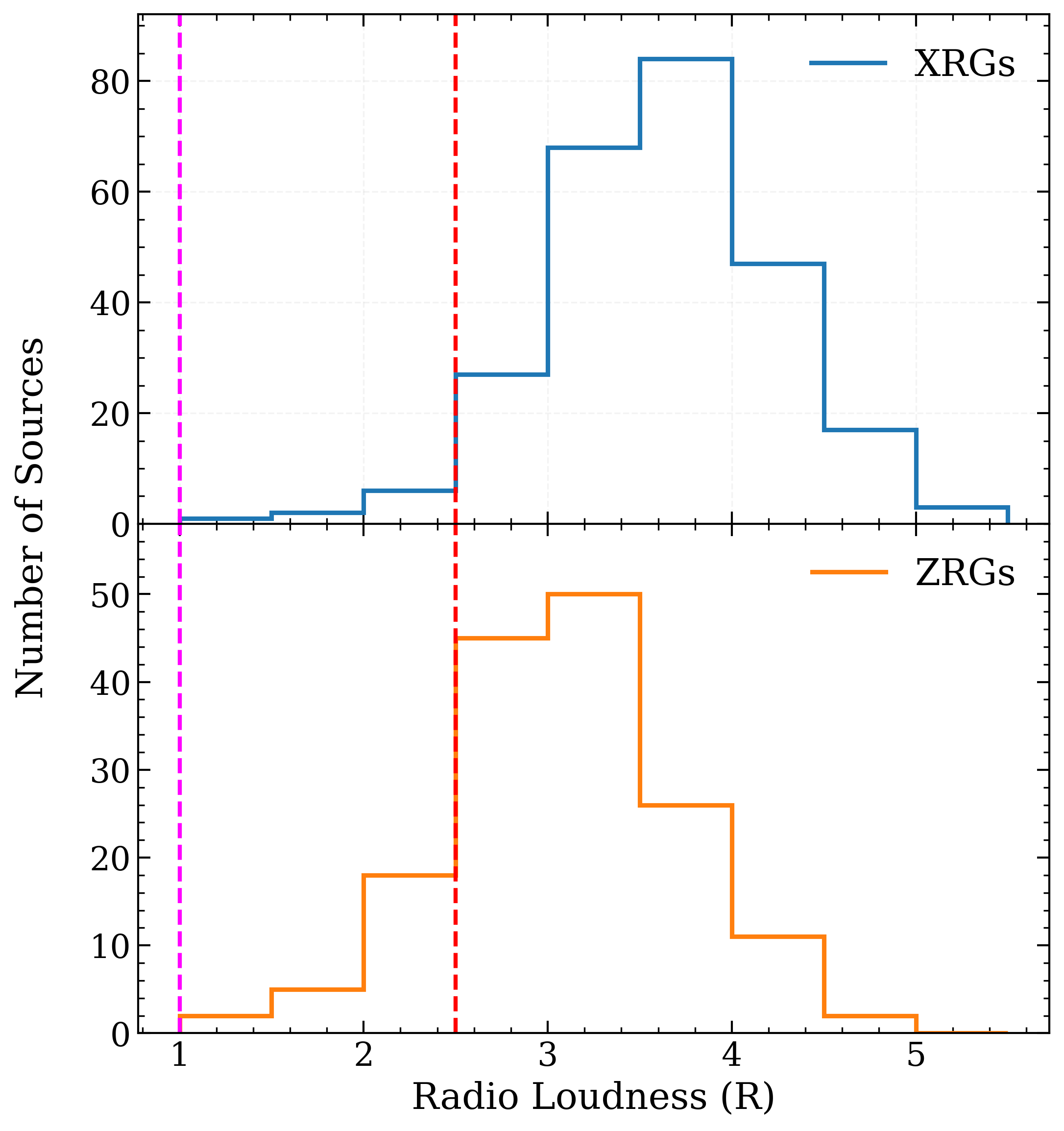}
\caption{Distribution of radio loudness parameter among the XRGs (upper panel) and ZRGs (lower panel). The vertical magenta and red dashed lines mark the boundaries at R = 1.0 and R = 2.5, are the thresholds for radio-loud and highly radio-loud as defined by \citep{2019MNRAS.482.2016Z} at 1.4 GHz; the thresholds are shown here for reference only.}
\label{fig:loudness}
\end{figure}

Radio loudness was estimated for 414 WRGs (255 XRGs and 159 ZRGs), spanning a range from 1.01 to 5.40, with a median value of 3.42 (Figure~\ref{fig:loudness}). The distribution indicates that WRGs are generally characterized by high values of the radio-to-optical loudness parameter, consistent with strongly radio-dominated systems.

For reference, \citet{2019MNRAS.482.2016Z} defined thresholds at 1.4 GHz ($R < 1.0$, $1.0 < R < 2.5$, and $R > 2.5$) corresponding to radio-quiet, radio-loud, and highly radio-loud regimes. While these thresholds cannot be directly applied to 144 MHz measurements, the observed distribution of $R$ values clearly places the WRG population in the regime of strong radio dominance. This is consistent with their jet-powered nature and predominantly FR II morphologies reported in \hyperlink{cite.2025ApJS..278...34B}{Paper I}.

A small subset of ten sources (3 XRGs and 7 ZRGs) shows relatively low radio loudness ($R < 2.0$). Based on the WISE color classification \citep{2014MNRAS.438..796S, 2015MNRAS.449.3191Y} (see Section~\ref{sec:infrared_prop}), nine of these objects are consistent with low-excitation radio galaxies (LERGs). The three sources with the lowest values ($R < 1.5$; J1237+4156, J1515+3116, and J1605+3150) have moderate 144 MHz flux densities (64–76 mJy) and are hosted by optically bright elliptical galaxies ($i \sim 14.5$–15.4), indicating that their low $R$ values primarily reflect relatively luminous hosts rather than intrinsically weak radio emission.

When divided into subclasses, XRGs are on average slightly more radio-loud than ZRGs, with mean (median) values of 3.63 (3.68) and 3.07 (3.05), respectively. Although this difference is modest, it is consistent with XRGs hosting somewhat more powerful or more efficiently collimated jets. However, the substantial overlap between the distributions indicates that radio loudness alone cannot account for the observed morphological differences between the two classes.

\section{Optical Properties}
\label{sec:optical_prop}

\begin{figure}[t!]
\includegraphics[angle=0,width=4.0cm,trim=3cm 0.5cm 3cm 0.5cm, clip]{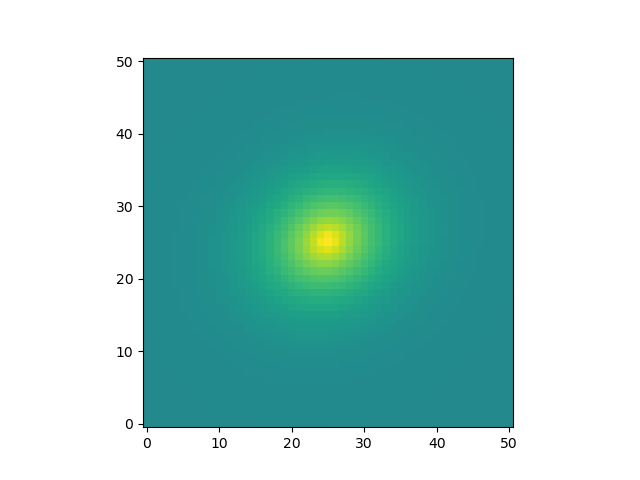}
\includegraphics[angle=0,width=4.2cm]{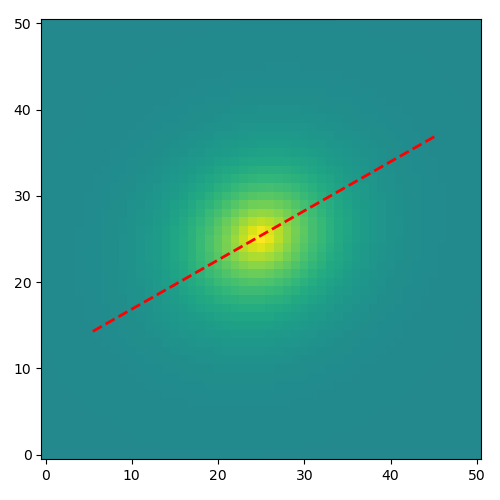}
\caption{Left: SDSS $i$-band image of SDSS J094438.52+641144.8, the optical host of the XRG J0944+6411. Right: the red dashed line marks the Sérsic-derived position angle of $120^\circ$. Both images cover 51×51 pixels (20 arcsec on a side) with a pixel scale of 0.396$^{\prime\prime}$ per pixel.}
\label{fig:sample_optical_PA}
\end{figure}

\subsection{Optical Position Angle and Ellipticity}
\label{sec:opa_ellips}
We measured the optical position angles (PAs) of the available host galaxies using SDSS $i$-band images, chosen for their higher signal-to-noise ratio (S/N). A two-dimensional Sérsic profile was fitted to each host \citep{2010AJ....139.2097P}, from which the optical PAs were derived. Reliable measurements were obtained for 432 sources. For the remaining objects, PAs could not be determined due to small host size, low surface brightness, or insufficient image quality, which hindered robust profile fitting. We also put a magnitude cut of $<$25 to remove the faint sources. An illustrative example of a host galaxy and its measured PA is shown in Figure \ref{fig:sample_optical_PA}. These optical PAs are subsequently used in the offset analysis (see Section \ref{sec:pa_offset}).

\begin{figure}[ht!]
\includegraphics[angle=0,width=8.50cm]{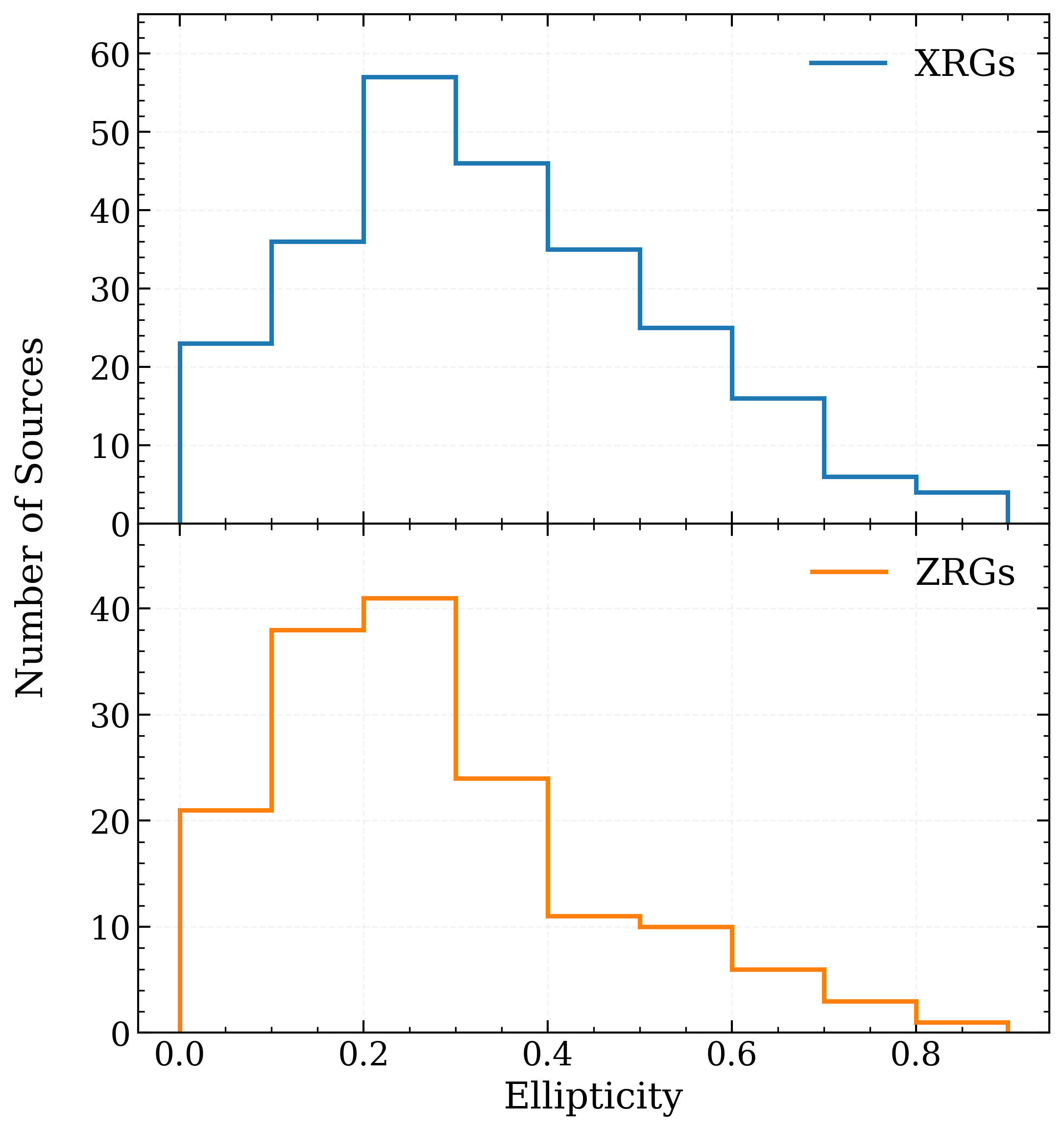}
\caption{Histogram of ellipticity for the host galaxies of XRGs (upper panel) and ZRGs (lower panel).}
\label{fig:ellipticity}
\end{figure}

We further examined the host ellipticities, defined as $\epsilon = 1 - b/a$, where $a$ and $b$ are the semi-major and semi-minor axes from the Sérsic fit. The ellipticity distribution is shown in Figure \ref{fig:ellipticity}, for the XRGs (upper panel) and ZRGs (lower panel). Across the full sample, the mean and median ellipticities are 0.31 and 0.29, respectively. When considered separately, XRG hosts exhibit higher ellipticities (mean 0.34, median 0.31) than ZRG hosts (mean 0.28, median 0.25). Although the absolute differences are modest, a Kolmogorov–Smirnov (KS) test indicates a statistically significant distinction between the two sub-populations ($D=0.183$, $p=0.003$). This result suggests that XRG and ZRG hosts may be intrinsically different, or at least possess systematically distinct structural properties.

\subsection{Radio-Optical Position Angle Offset}
\label{sec:pa_offset}
To examine the geometrical connection between radio morphology and either the host galaxy structure or the surrounding environment, we measured angular offsets using the position angles derived in Sections~\ref{subsec:radio_pa} and~\ref{sec:opa_ellips}. Given the intrinsic morphological differences between XRGs and ZRGs, we adopt distinct definitions of the relevant offset angles for the two classes, following the methodology described in Section~\ref{subsec:radio_pa}. This distinction is physically motivated and reflects the different spatial origin and nature of the wings in the two morphologies.

\begin{figure}[ht!]
\includegraphics[angle=0,width=8.50cm]{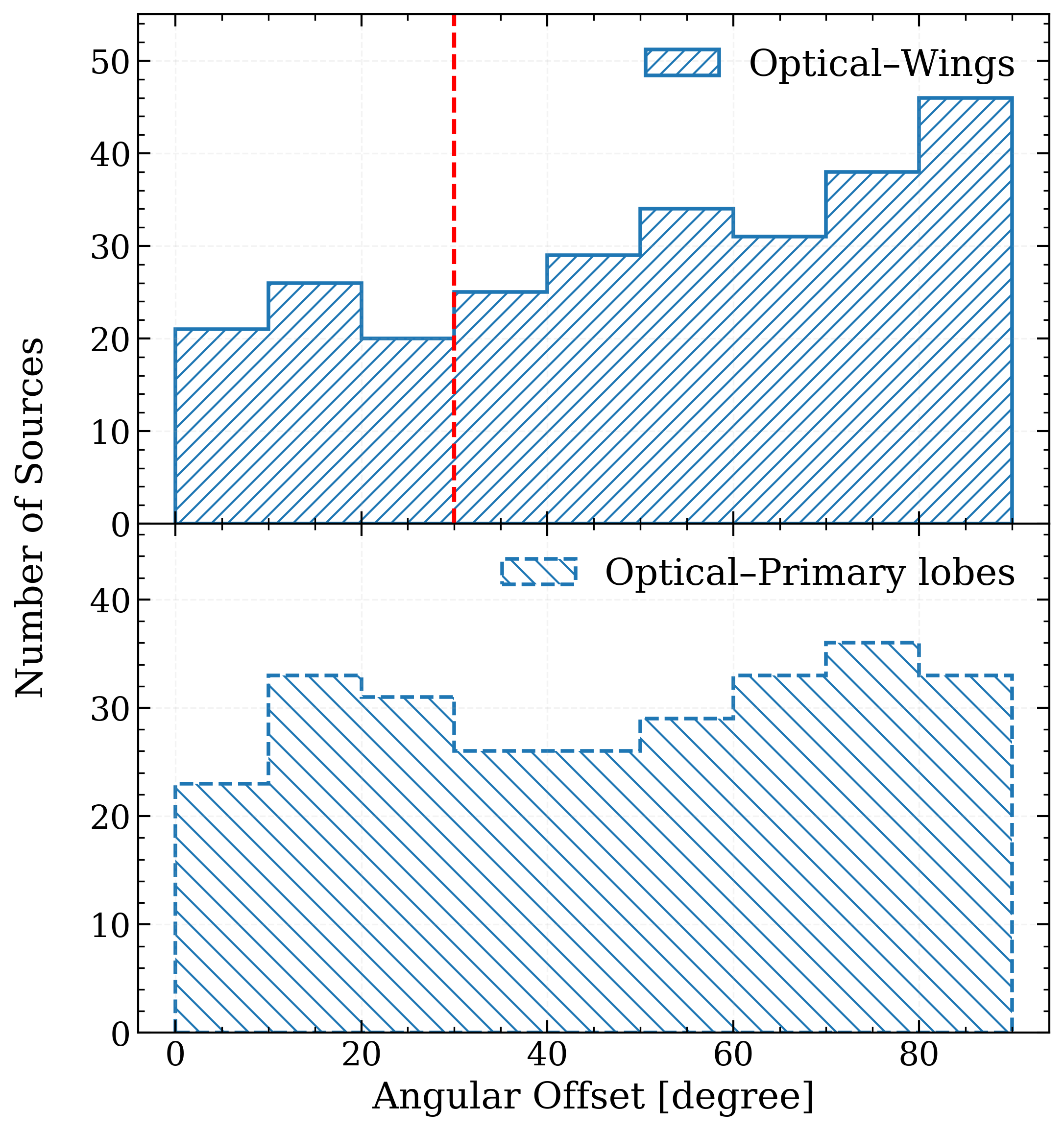}
\caption{Distributions of angular offsets for the XRGs in our sample. The upper panel shows the position angle offset between the optical major axis of the elliptical host and radio wings, whereas the lower panel presents the offset between the optical axis and primary radio lobes. The vertical red dashed line in the upper panel marks the 30$^\circ$ offset.}
\label{fig:radio_optical_offset_xrg}
\end{figure}

For the XRGs, we calculated the angular offsets between the optical major axis of the host galaxy and both the radio wings and the primary radio lobes. The wing–optical offset metric directly probes the hydrodynamic backflow scenario, in which synchrotron plasma from the primary lobes flows back toward the host and is redirected along the steepest pressure gradient in an anisotropic gaseous halo, typically aligned with the minor axis of an elliptical galaxy (i.e. $\sim$90$^\circ$ from the optical major axis). The resulting wing–optical and optical–primary lobe offset distribution for the XRGs is shown in Figure~\ref{fig:radio_optical_offset_xrg}. For each source, we used the mean of the two wing offsets and considered the acute angle.

The XRG population, comprising 270 sources, shows a strong preference for large misalignments between the optical major axis and the radio wings. In particular, 75\% of the sample exhibits wing–optical offsets greater than 30$^\circ$, with a median value of 54$^\circ$, and 17\% display nearly perpendicular configurations ($\geq$80$^\circ$). A Kolmogorov–Smirnov test strongly rejects the null hypothesis of a uniform offset distribution, yielding a p-value of $\sim7\times10^{-4}$. These results are fully consistent with previous studies based on smaller samples selected from shallower or higher-frequency radio surveys. Early work by \citet{2002A&A...394...39C} on a sample of nine XRGs reported a strong tendency for wings to align with the host minor axis, a result later confirmed by \citet{2009ApJ...695..156S}. The most systematic pre-LoTSS study was carried out by \citet{2016A&A...587A..25G}, who analysed 22 well-defined XRGs selected from FIRST and found a near-absence of sources with wing–optical offsets below 30$^\circ$. More recently, \citet{2019ApJ...887..266J}, using 41 FIRST XRGs, also reported a preference for large misalignments, although about 24\% of their sources exhibited offsets below 30$^\circ$.

In contrast, the optical–primary lobe distribution does not show any clear preference for either alignment or perpendicularity, and is consistent with a uniform distribution (KS test, $p=0.45$), in agreement with the results of \citet{2019ApJ...887..266J}. 
This lack of correlation indicates that the orientation of the primary jets is not strongly linked to the large-scale stellar structure of the host galaxy. Consequently, the pronounced alignment observed for the wings is unlikely to arise from a direct coupling between the jet axis and the host galaxy geometry. Instead, it supports a scenario in which the wings are shaped by backflow within an anisotropic gaseous halo, whose structure is influenced by the overall mass distribution of the galaxy.

Our XRG sample, drawn from the deeper and higher-fidelity LoTSS DR2 data, is substantially larger than those considered in earlier studies and confirms these trends with higher statistical significance (Figure~\ref{fig:radio_optical_offset_xrg}). At the same time, the increased sensitivity reveals a non-negligible minority of sources (25\%) with small wing–optical offsets ($<$30$^\circ$), including $\sim$8\% with near alignment ($<$10$^\circ$). Such systems were rare or absent in earlier, more strictly selected samples, indicating that selection effects and data quality play an important role in shaping the observed offset distributions. While the backflow model provides a natural explanation for the majority of XRGs, these exceptions suggest that additional mechanisms—such as jet reorientation, episodic activity, or merger-driven changes in jet direction—may also contribute to the formation of wings in a subset of sources \citep[e.g.][]{2023MNRAS.523.1648M, 2025MNRAS.536.2025M, 2024ApJ...969..156S}. This supports a picture in which XRGs are not produced by a single universal mechanism.

For the ZRGs, the situation is fundamentally different. In these sources, the wings do not emanate from the vicinity of the SMBH and therefore do not define a global morphological axis related to the stellar potential of the host galaxy. As a result, offsets between the wings and the optical major or minor axis are not physically informative and do not provide a meaningful test of any specific formation model. Instead, for the ZRG subsample we quantify the angular offset between the wings and the primary radio lobe axis, which characterizes the degree of bending or deformation of the radio structure on kiloparsec scales. This wing–lobe offset measures how strongly the secondary structures deviate from the original jet direction and is sensitive to the combined effects of jet stability, intrinsic changes in jet direction, and interactions with an asymmetric external medium, particularly near the lobe termini.

\begin{figure}[ht!]
\includegraphics[angle=0,width=8.50cm]{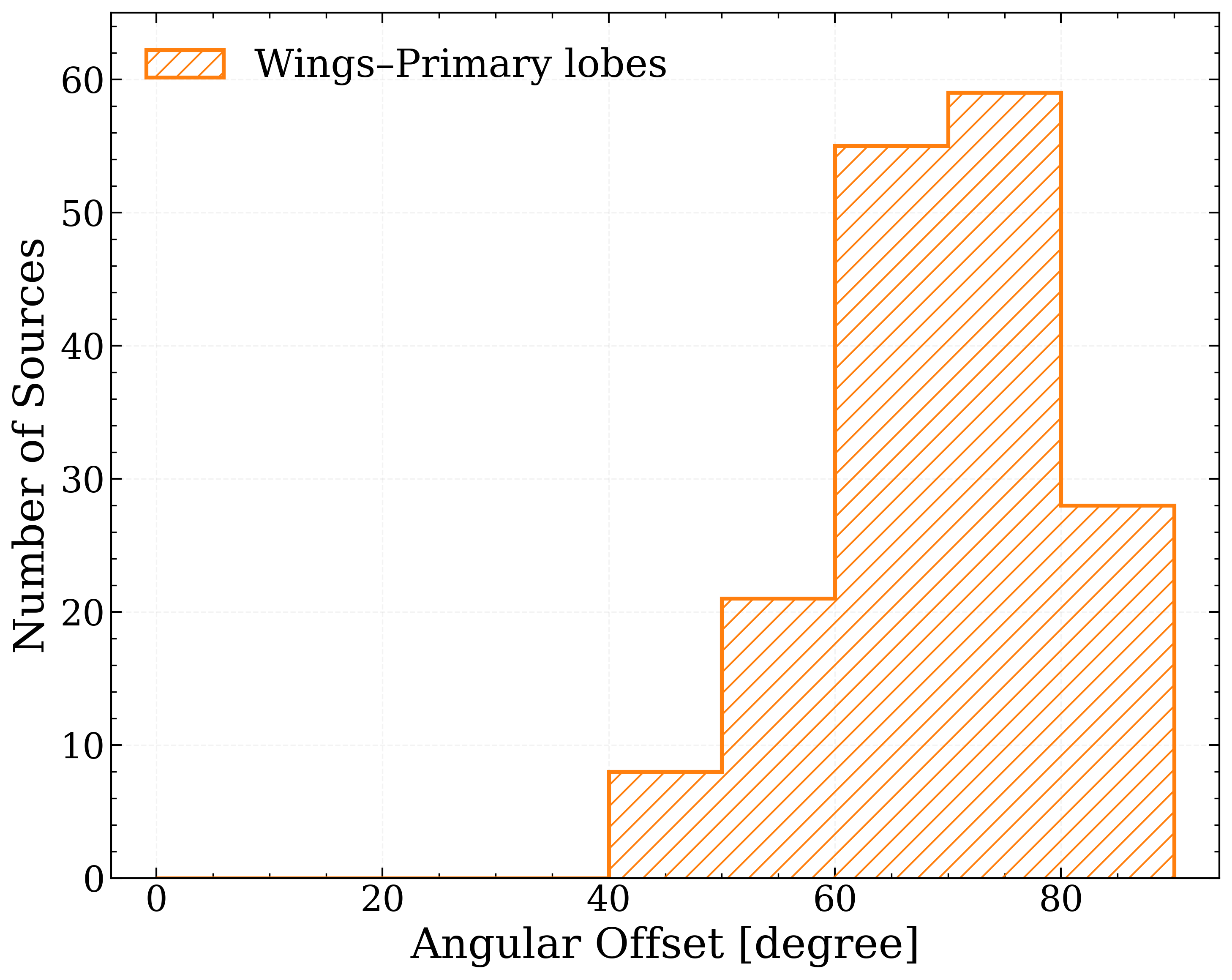}
\caption{Distributions of angular offsets for the ZRGs in our sample. Here, the angular offset is measured between the wing axis and the primary radio lobe axis.}
\label{fig:offset_zrg}
\end{figure}

The resulting wing–lobe offset distribution for the ZRGs (171 sources; Figure~\ref{fig:offset_zrg}) is strongly skewed toward large angles. The majority of ZRGs (83\%) exhibit offsets exceeding 60$^\circ$, with a broad maximum between 60$^\circ$ and 80$^\circ$, and no sources are found with offsets below 40$^\circ$. At face value, this indicates that ZRG wings are not minor perturbations of the jet axis but instead represent strongly deflected or displaced structures. However, it is important to note that this behaviour is, at least in part, intrinsic to the morphological selection of ZRGs: by definition, Z-shaped sources are identified by pronounced lateral displacements of the wings relative to the lobes, which naturally biases the sample toward large wing–lobe offsets. Consequently, the observed distribution should not be interpreted as evidence for a preferred physical angle or a unique formation mechanism.

Nevertheless, the predominance of large wing–lobe offsets in ZRGs, together with the near absence of small deflections, provides important constraints on the physical mechanisms shaping their morphology. While such large offsets are often interpreted as signatures of strong jet–environment interactions, this explanation alone is insufficient to account for the characteristic Z-shaped geometry. In particular, purely local or stochastic deflections would more naturally produce one-sided or co-aligned distortions, rather than the point-symmetric configuration in which wings emerge from opposite lobe termini in opposing directions. The observed morphology therefore points to a mechanism that acts coherently on both jets, with the external medium likely playing a secondary, modulatory role.

Results from our previous study (\hyperlink{cite.2025ApJS..278...34B}{Paper I}) show that the radio power distributions of winged sources at both 1.4 GHz and 144 MHz are largely indistinguishable from those of regular FR-II radio galaxies, with XRGs being systematically more luminous than ZRGs. Moreover, the vast majority of winged sources ($\sim$88\%) exhibit FR-II morphologies, a fraction that increases to $\sim$95\% for XRGs alone. In contrast, the FR-I population is dominated by ZRGs, while FR-I XRGs are almost entirely absent. This suggests that edge-darkened winged sources preferentially develop Z-shaped rather than X-shaped morphologies, consistent with lower jet power and reduced collimation making the jets more susceptible to large-scale deformation.

Within this framework, the large wing–lobe offsets measured in ZRGs are consistent with scenarios in which jets are more easily perturbed near the lobe termini, either through interaction with an asymmetric external medium or through intrinsic changes in jet direction, such as slow reorientation or episodic activity. Importantly, the systematic antisymmetry of the Z-shaped structure disfavors purely local deflections as the primary driver and instead suggests that environmental effects act to shape or amplify an underlying global deformation of the jet system.

\begin{figure*}[ht!]
\includegraphics[angle=0,width=9.0cm]{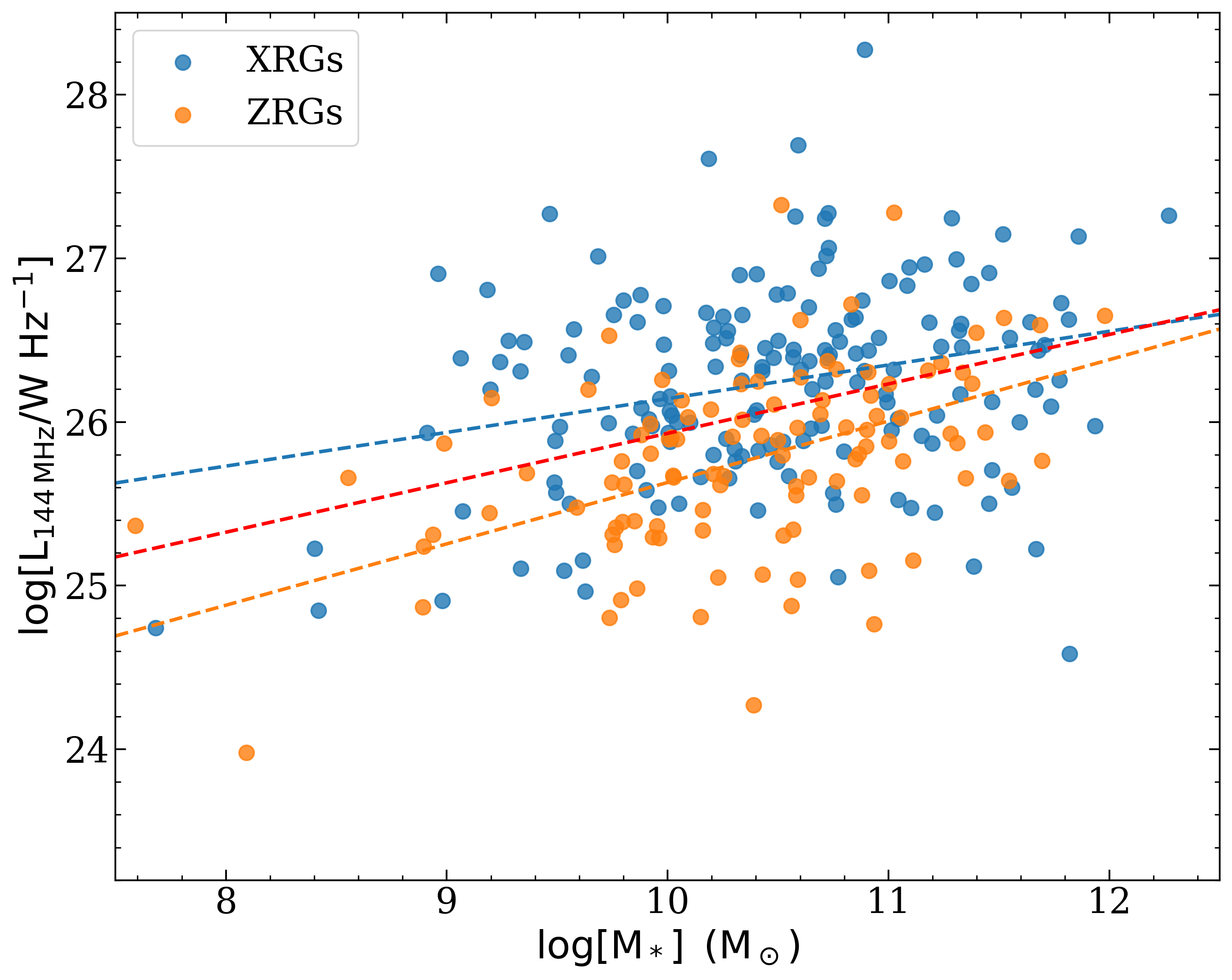}
\includegraphics[angle=0,width=9.0cm]{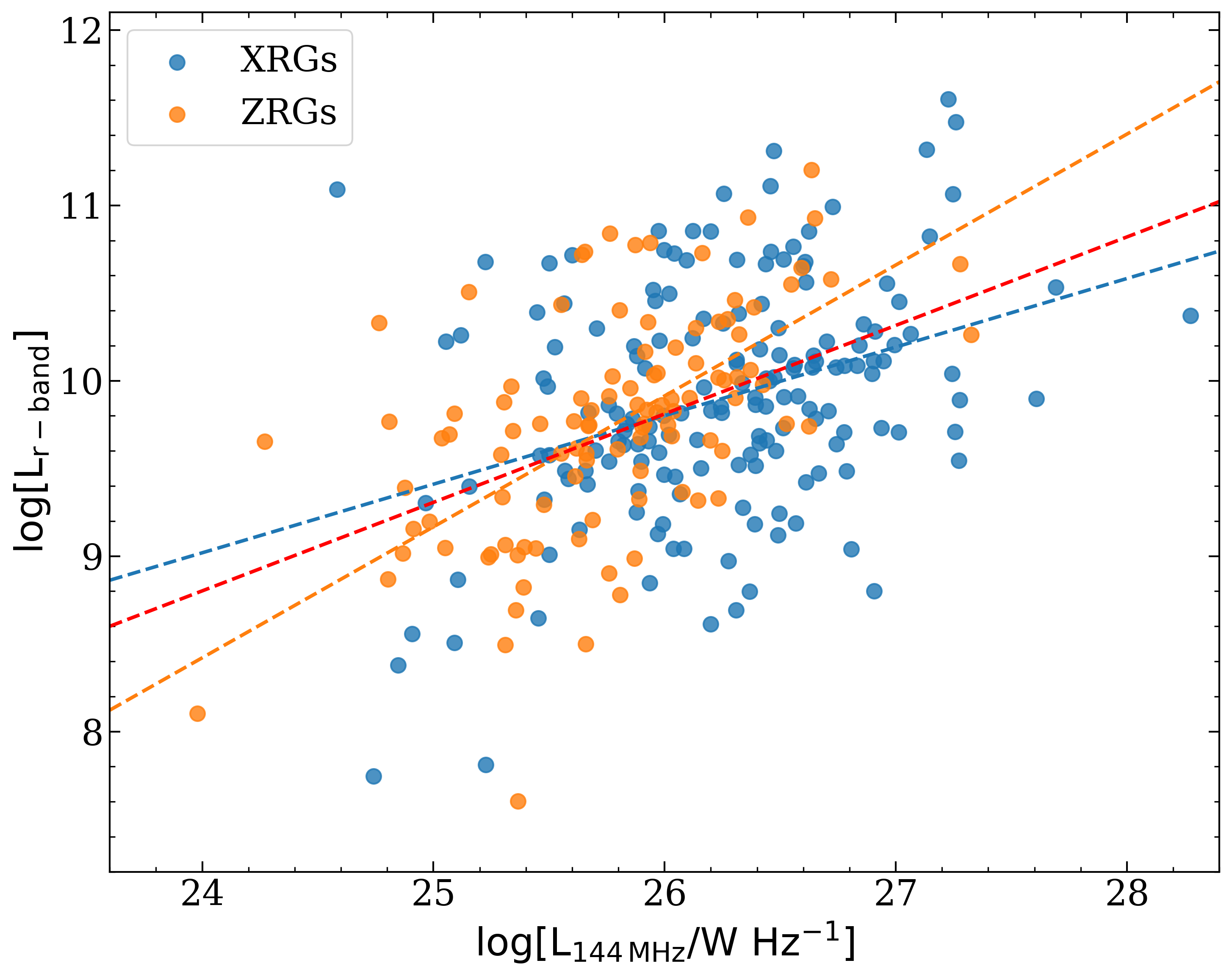}
\caption{Left panel: The relation between the stellar mass ($M_*$) of the host and the LoTSSS 144 MHz radio luminosity ($L_{144MHz}$) for the WRGs. Right panel: The 144 MHz radio luminosity from LoTSS ($L_{144MHz}$) vs. r-band luminosity ($L_{r-band}$). In both pictures, the dashed line represents the fitted trends for the XRGs (blue line), ZRGs (orange line), and all of the WRGs (red line).}
\label{fig:mass_map}
\end{figure*}

\subsection{Host Stellar Mass}
\label{subsec:stellar_mass}
We estimated host stellar masses ($\rm M_*$) using optical photometry from SDSS DR18. Stellar masses were derived with the empirical color-based mass-to-light ratio method of \citet{2003ApJS..149..289B}, adopting the $r$-band relation and a Chabrier initial mass function (IMF) \citep{2003PASP..115..763C, 10.1093/mnras/staf622}:

\begin{equation}
\log_{10}(M_{*}/L_{r-band}) = -0.456 + 1.097 \times (g - r),
\end{equation}
where $\rm M_*/L_{r-band}$ is the stellar mass-to-light ratio in solar units and $g-r$ is the rest-frame color (AB magnitudes). The zero point ($-0.456$) includes a $-0.15$ dex correction relative to a diet Salpeter IMF, consistent with a Chabrier IMF. The $r$-band luminosity was then computed as $\rm L_{r-band} = 10^{-0.4(M_r - M_{r,\odot})}$, adopting $\rm M_{r,\odot} = 4.65$ in the AB system \citep{2018ApJS..236...47W}.

We restricted the sample to sources with $0.3 \leq g-r \leq 1.5$, encompassing typical early-type galaxies (0.6–1.0), slightly earlier spirals (0.3–0.6), and red ellipticals with dust or metallicity effects (up to 1.5) \citep{2005ApJ...629..143B, 2007AJ....134..579F}. Objects outside this range were excluded due to possible dust reddening, photometric errors, or other uncertainties. This selection yielded 276 sources with reliable stellar mass estimates, among which 169 are XRGs, and 107 are ZRGs.

The derived stellar masses span $\log(M_*/M_\odot) = 7.59$–12.56, with a median value of 10.49. The $r$-band luminosities have mean and median values of 9.84 and 9.83, respectively, in units of $\log(L_{r-band}/L_\odot)$. Figure~\ref{fig:mass_map} shows the stellar mass ($M_*$) and $r$-band luminosity ($L_{r-band}$) as a function of the 144 MHz radio luminosity ($L_{144\,MHz}$) from LoTSS DR2. When considered separately, XRGs have median stellar masses and luminosities of $\log(M_*/M_\odot) = 10.52$ and $\log(L_{r-band}/L_\odot) = 9.89$, respectively, whereas ZRGs exhibit slightly lower median stellar masses ($10.41$) and $r$-band luminosities ($9.76$).

Taken together, these results indicate that both XRGs and ZRGs are predominantly hosted by massive and luminous galaxies, consistent with the elliptical-dominated host populations reported in previous LoTSS studies \citep{2023A&A...678A.151H}. 
The close similarity in the stellar mass and luminosity distributions of the two subclasses further suggests that global host-galaxy properties alone are insufficient to explain the observed morphological differences.

\subsection{Environment Study}
\label{subsec:enviroment}
To investigate the environments of WRGs, we quantified the local galaxy density around each source. Following the method of \citet{2019ApJ...887..266J}, we counted photometric galaxies within a projected radius of 1.0 Mpc and within a photometric redshift slice of $\pm 0.04 \times (1+z)$ centered on the host-galaxy redshift, using the SDSS DR18 photometric catalog. Because photometric redshift uncertainties increase with redshift and incompleteness becomes significant at faint magnitudes, we restricted our analysis to sources with $z \leq 0.6$ (median $z \approx 0.52$). To further mitigate redshift-dependent incompleteness, only galaxies with $r$-band model magnitudes $< 21$ and absolute magnitudes $M_r < -19$ were included.

\begin{figure}[ht!]
\includegraphics[angle=0,width=8.5cm]{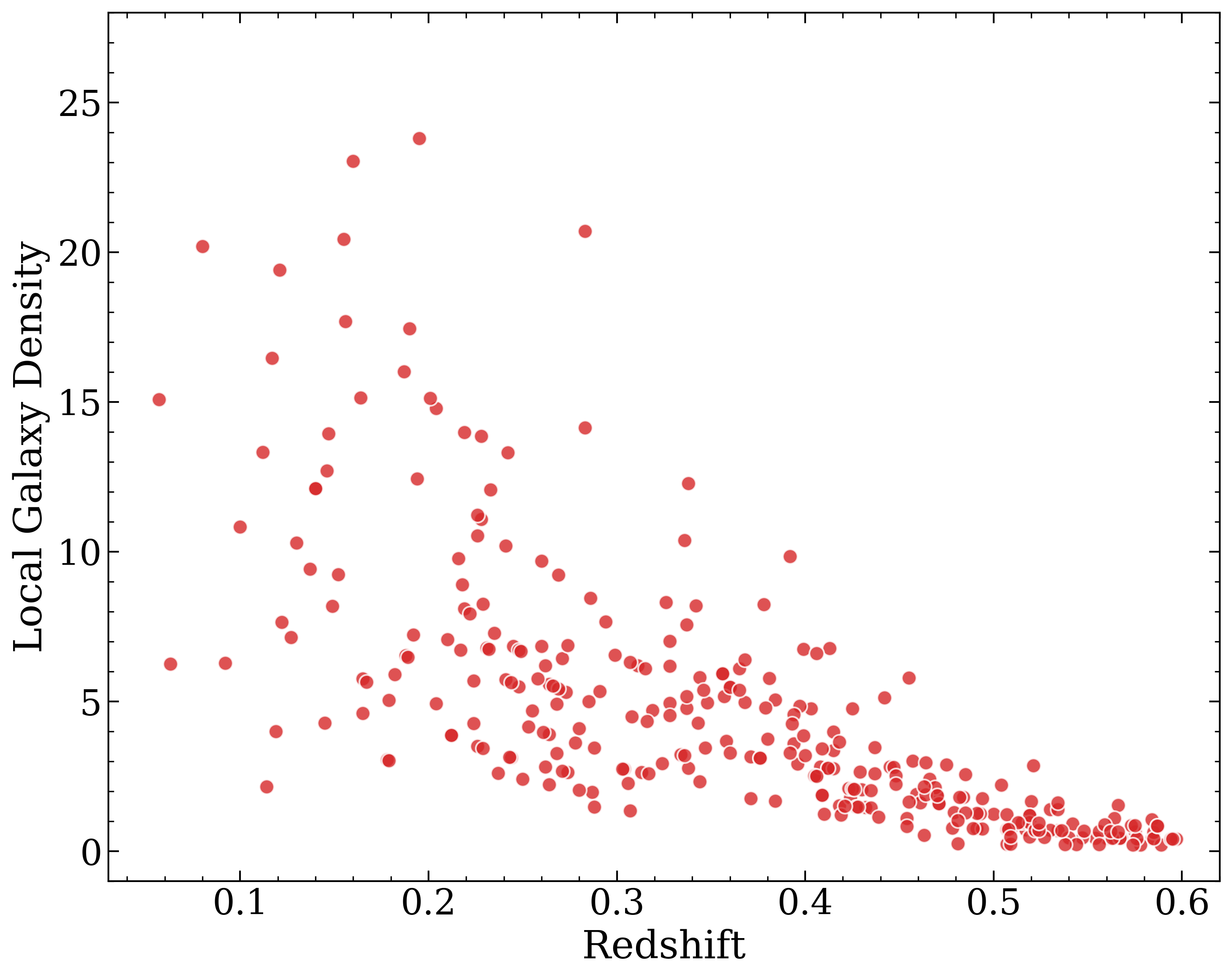}
\caption{Redshift versus local galaxy density for the WRG sample. The distribution shows a systematic decrease in density with increasing redshift.}
\label{fig:galx_denst}
\end{figure}

Applying these constraints yielded reliable local density measurements for 316 WRGs (192 XRGs and 124 ZRGs). The mean and median local galaxy densities for the full sample are 5.5 and 4.3 galaxies Mpc$^{-2}$, respectively. Separately, XRGs have slightly higher mean and median densities (5.7 and 4.8) than ZRGs (5.4 and 4.1), though the difference is marginal. The redshift–density distribution is shown in Figure~\ref{fig:galx_denst}. At low redshifts, local densities are often higher and exhibit considerable scatter. With increasing redshift, the measured densities decline, reflecting both the loss of fainter galaxies beyond the SDSS detection limit (under the $M_r < -19$ cut) and the growing uncertainties in photometric redshifts, which dilute associations within the adopted slice.

We also examined cluster associations by cross-matching WRG positions with the \citet{2012ApJS..199...34W} galaxy cluster catalog. A cluster was considered associated if it lay within 1.5 Mpc projected distance and within $\pm 0.04 \times (1+z)$ of the WRG redshift. This search identified 51 WRGs with potential cluster associations, including 25 XRGs and 26 ZRGs.

Overall, these results indicate that both XRGs and ZRGs typically reside in environments of moderate galaxy density. Although XRGs exhibit slightly higher average local densities, the difference with respect to ZRGs is small. The comparable fractions of cluster-associated sources in both subclasses further suggest that the large-scale environment is not the primary driver of the morphological distinction between XRGs and ZRGs, even if ZRGs may be marginally more prevalent in richer cluster environments.

\section{Infrared Properties}
\label{sec:infrared_prop}
\begin{figure}[t!]
\includegraphics[angle=0,width=8.50cm]{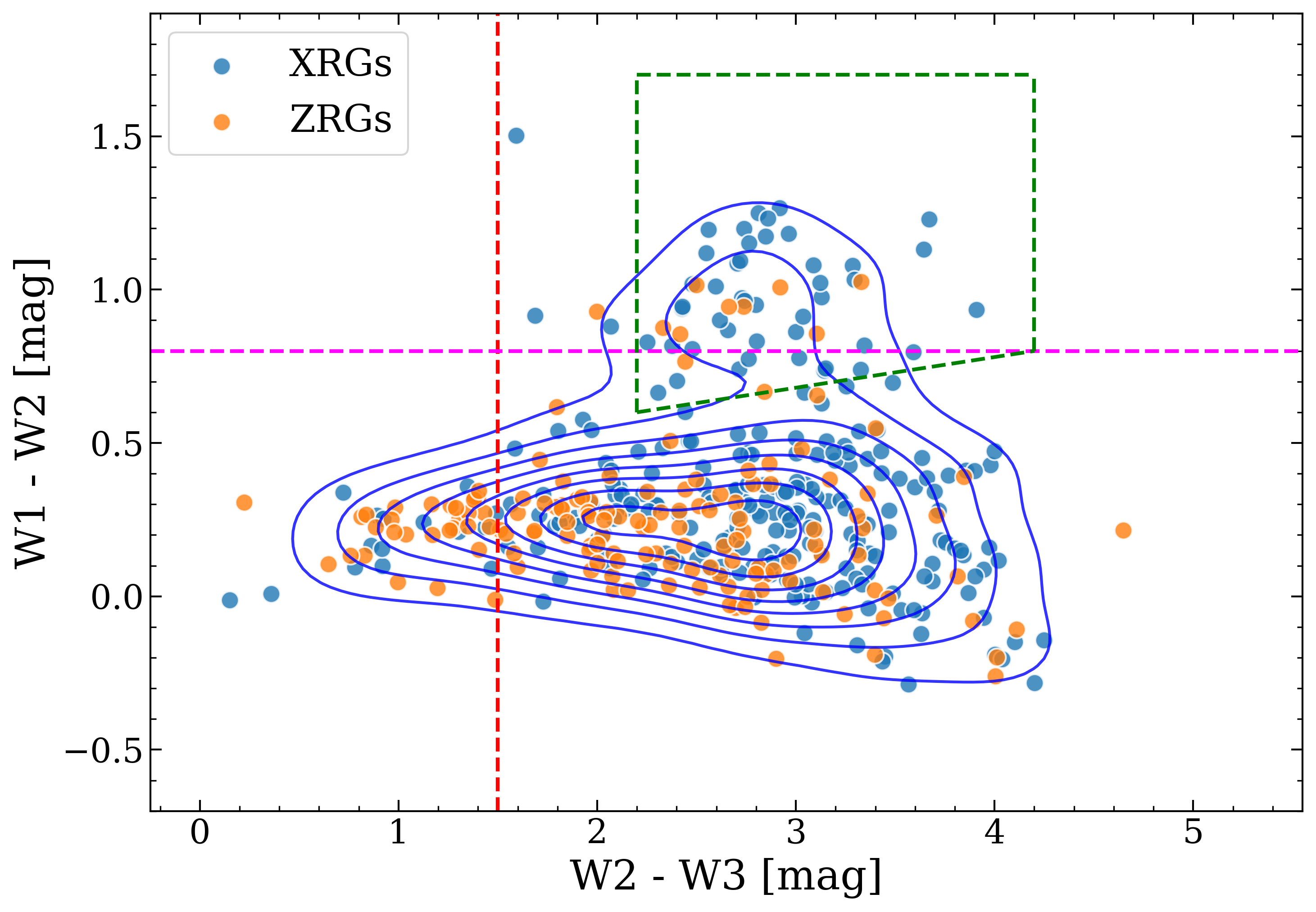}
\caption{WISE color–color diagram for the WRGs. The vertical red dashed line at W2–W3 = 1.5 separates typical ellipticals (left) and spirals (right) \citep{2010AJ....140.1868W}. The magenta dashed line at W1–W2 = 0.8 marks the threshold for WISE QSOs/powerful AGN \citep{2012ApJ...753...30S}. The green quadrilateral indicates the “AGN box” \citep{2011ApJ...735..112J}.}
\label{fig:wise_map}
\end{figure}

We investigated the mid-infrared properties of the WRG sample using data from the Wide-field Infrared Survey Explorer \citep[WISE;][]{2010AJ....140.1868W} all-sky survey \citep[AllWISE;][]{2014yCat.2328....0C}. 
WISE counterparts to the optical hosts galaxies were identified within a maximum separation of 3 arcsec from the SDSS positions \citep{2019ApJ...887..266J}, yielding detections for 409 sources, including 258 XRGs and 151 ZRGs. Photometry in the W1 ($\rm 3.4 \mu m$), W2 ($\rm 4.6 \mu m$), and W3 ($\rm 12 \mu m$) bands was extracted from the IRSA Gator catalog\footnote{\url{https://irsa.ipac.caltech.edu/applications/Gator/}} and used to compute the infrared colors W1–W2 and W2–W3.
The resulting color–color distribution is shown in Figure~\ref{fig:wise_map}.

According to \citet{2010AJ....140.1868W}, galaxies with W2–W3 $>$ 1.5 
typically exhibit mid-infrared colors associated with enhanced dust emission and cold interstellar medium, while those with W2–W3 $<$ 1.5 
are more consistent with dust-poor, quiescent systems.
In our sample, 89\% of WRGs lie in the W2–W3 $>$ 1.5 region, indicating that the majority of sources possess significant amounts of dust and/or cold gas. This trend is particularly pronounced for XRGs, where 94\% fall in this region, compared to 80\% for ZRGs. Such infrared properties may reflect the presence of residual ISM or recent gas-rich interactions, despite the predominantly elliptical optical morphologies of the hosts.

We further applied established mid-IR AGN selection criteria. Using the conservative cut W1–W2 $\geq 0.8$ \citep{2012ApJ...753...30S}, 11\% of WRGs are classified as hosting powerful AGN (WISE-selected QSOs). Adopting a less stringent cut of W1–W2 $\geq 0.6$ \citep{2010AJ....140.1868W}, increases this fraction to 16\%. In both cases, XRGs exhibit systematically higher AGN fractions (15–20\%) than ZRGs (6–9\%). Consistently, 12\% of WRGs fall within the WISE-defined “AGN box” \citep{2011ApJ...735..112J}, including 16\% of XRGs and only 6\% of ZRGs. These results indicate that XRGs are more frequently associated with radiatively efficient, powerful AGN activity than ZRGs.

Using the WISE diagnostic criteria of \citet{2014MNRAS.438.1149G}, we further classified WRGs into low-excitation radio galaxies and high-excitation radio galaxies (HERGs). Sources with W2–W3 $<$ 2 and W1–W2 $<$ 0.4 are considered LERGs \citep{2014MNRAS.438..796S, 2015MNRAS.449.3191Y}. Based on this classification, 21\% of WRGs in our sample are consistent with LERGs, in good agreement with the fractions reported by \citet{2016A&A...587A..25G, 2019ApJ...887..266J}. When considering the subclasses separately, 12\% of XRGs and 36\% of ZRGs fall into the LERG category.

Taken together, these infrared diagnostics show that although WRGs are predominantly hosted by massive elliptical galaxies (Section~\ref{subsec:stellar_mass}), their mid-infrared properties frequently resemble those of spirals, indicating the presence of dust and cold gas. This suggests that merger-driven processes, residual ISM, or ongoing gas accretion may be common in these systems. The higher AGN fraction observed among XRGs points to a closer connection with efficient nuclear accretion, whereas the higher LERG fraction among ZRGs is consistent with lower radiative efficiency and more quiescent accretion modes, often associated with denser, cluster-like environments (see Section~\ref{subsec:enviroment}).

It is important to note that the apparent discrepancy between the optical and infrared diagnostics is only superficial. While optical luminosities and stellar masses clearly identify WRG hosts as massive ellipticals (Section~\ref{subsec:stellar_mass}), their mid-infrared colors often fall in the “spiral” region of the WISE diagram. This apparent inconsistency can be reconciled by considering that mid-infrared colors are sensitive to dust and cold gas, rather than stellar morphology. Consequently, WRGs can retain elliptical morphologies while simultaneously exhibiting infrared signatures of enhanced ISM content or star formation. This highlights that large-scale host galaxy properties alone cannot explain the observed diversity among WRGs, and that merger-driven processes and AGN fueling play a key role in shaping their radio morphologies and evolutionary pathways.

\section{X-ray properties}
\label{sec:xray_prop}
Several observational studies have identified X-ray cavity systems associated with XRGs, such as 4C +00.58 and NGC 326 \citep{2010ApJ...717L..37H, 2012ApJ...746..167H}. In addition, X-ray cavities aligned nearly perpendicular to the primary jets have been reported in systems like NGC 193 and RBS 797 \citep{2014ApJ...782L..19B, 2021ApJ...923L..25U}. Numerical simulations further suggest that such structures could arise from jet reorientation events \citep{2018arXiv180104276C, 2022ApJ...936L...5L, 2023ApJS..268...49G}, indicating that X-ray observations can provide insight into WRG formation in the context of backflow and jet-reorientation models.

We searched for X-ray counterparts to our WRG sample using archival catalogs accessible via the HEASARC interface\footnote{\url{https://heasarc.gsfc.nasa.gov/xamin/}}. The primary catalogs included the second ROSAT All-Sky Survey \citep[2RXS;][]{2016A&A...588A.103B}, the XMM-Newton Serendipitous Source Catalog \citep[XMMSSC;][]{2020A&A...641A.136W}, the Chandra Source Catalog release 2.1.1 \citep[CSC v2.1.1;][]{2024ApJS..274...22E}, and the eROSITA All-sky Survey first 6 months X-ray source catalog (0.2$-$2.3 keV) \citep[eRASS1main;][]{2024A&A...682A..34M}. A positional cross-match with a radius of 30 arcseconds was performed. Using these catalogs, only 28 WRGs ($\sim$ 6\% of the sample) were found to have potential X-ray counterparts: 14 from 2RXS, 4 from CSC v2.1.1, 6 from XMMSSC-DR14, and 4 from eRASS1. Sources with these possible X-ray detections are given in the Appendix section \ref{appendix_xray_table}.

For these detected sources, we checked the catalogued X-ray parameters. The 2RXS sources exhibit count rates in the range 0.013--0.11 ct s$^{-1}$, with a number of objects (five) showing non-zero extent probabilities, and these might be some group or cluster-scale hot gas. The sources detected from Chandra have 0.5--7.0 keV band fluxes from 10$^{-14}$ to 10$^{-12}$ erg s$^{-1}$ cm$^{-2}$, indicating both possible faint nuclear emission and bright AGN. The four eROSITA detections have 0.2--0.6 keV fluxes of $\sim$(0.7--2.3) $\times$ 10$^{-13}$ erg s$^{-1}$ cm$^{-2}$, with a count rate of $\sim$ 0.08 to 0.24 ct s$^{-1}$. The XMM-Newton detections cover a range from 10$^{-15}$ to 10$^{-13}$ erg s$^{-1}$ cm$^{-2}$ in the 0.2--12.0 keV band, indicating both faint and moderately bright X-ray sources.

The low detection fraction motivated a preliminary photometric analysis using the full eRASS data; however, no fluxes were measured above the 1$\sigma$ level. This limited detection is likely due to factors including the incomplete and heterogeneous sky coverage of the X-ray surveys relative to LoTSS DR2, as well as the relatively shallow sensitivity of current X-ray data, which hampers the detection of faint sources. Consequently, the paucity of X-ray counterparts prevents robust correlation studies between X-ray and radio properties for our WRG sample.

Nevertheless, the few detections, in line with previous X-ray observations of similar sources, indicate that a small subset of WRGs is associated with X-ray-emitting AGN or clusters. Future wide-area and deeper surveys, such as the forthcoming full-area eROSITA data release, or targeted X-ray observations, will be essential to better characterize the X-ray properties of WRGs and their connection to radio morphology.

\section{Summary}
\label{sec:summary}
Winged radio galaxies (WRGs) constitute a rare class of radio-loud AGN whose complex radio morphologies offer valuable constraints on jet physics, host galaxy structure, and environmental effects. Using a large sample selected from the LoTSS DR2 survey, we conducted a multiwavelength analysis to investigate the origin of X-shaped (XRGs) and Z-shaped (ZRGs) radio morphologies. Our main results can be summarized as follows:

\begin{itemize}
    \item[-] WRGs are predominantly strongly radio-dominated systems, with the vast majority exhibiting radio loudness values characteristic of powerful jet-dominated sources. XRGs are, on average, more radio-loud than ZRGs and are strongly dominated by FR II morphologies, whereas ZRGs more frequently include FR I sources.

    \item[-] XRGs show a strong preference for large misalignments between the radio wings and the optical major axis, consistent with hydrodynamic backflow along the minor axis of anisotropic gaseous halos. However, a substantial fraction of XRGs exhibits small or near-zero offsets, indicating that backflow alone cannot account for all cases and that additional processes, such as jet reorientation or episodic activity, likely contribute in a subset of sources.
    In contrast, ZRGs are characterized by predominantly large wing–lobe offsets, reflecting a high degree of large-scale deformation of the radio structure. Rather than arising from purely stochastic local deflections, these offsets likely result from a combination of reduced jet stability, intrinsic changes in jet direction, and interactions with an asymmetric external medium, acting coherently on kiloparsec scales near the lobe termini.

    \item[-] Host galaxy properties show that both XRGs and ZRGs reside in massive, luminous galaxies with broadly similar stellar masses and optical luminosities, indicating that global host properties alone do not drive the observed morphological differences. Environmental analysis further shows that WRGs typically inhabit low- to moderate-density environments, with only a small fraction associated with rich clusters. ZRGs exhibit a slightly higher incidence of cluster environments, suggesting that denser media may facilitate the development of Z-shaped morphologies through enhanced jet bending or confinement.

    \item[-] The majority of WRGs display mid-infrared colors indicative of substantial dust and cold gas content, despite their elliptical hosts. XRGs are more frequently associated with powerful AGN signatures, while ZRGs show a higher fraction of low-excitation radio galaxies. This points to more efficient accretion and higher jet power in XRGs, and comparatively radiatively inefficient, more quiescent systems among ZRGs.

    \item[-] X-ray counterparts are detected only for a small fraction of the sample, likely reflecting survey sensitivity and coverage limitations, but indicating that some WRGs host X-ray-emitting AGN or reside in hot gaseous environments.
    
\end{itemize}

Overall, our results indicate that the diversity of winged radio galaxy morphologies cannot be explained by a single formation pathway. Instead, their properties are best understood within hybrid scenarios in which jet power and stability, accretion mode, host galaxy structure, and environment act together to shape the observed radio morphologies. Deeper multiwavelength data and larger samples will be essential to further disentangle the relative importance of these processes.

\section*{Acknowledgments}
We are thankful to the anonymous referee for the thoughtful and constructive suggestions that helped to improve this paper. This work is supported by the National SKA Program of China No. 2025SKA0150103, National Natural Science Foundation of China under Nos. 12550002, 12133008, 12221003, 11890692. S.K.B. and T.F. acknowledge the science research grants from the China Manned Space Project with No. CMS-CSST-2021-A04 and No. CMS-CSST-2025-A10. T.K.S. and X.C. acknowledge funding support from the National SKA Program of China (Nos. 2022SKA0110100 and 2022SKA0110101). SYU acknowledges support from the UTokyo Global Activity Support Program for Young Researchers.

This research has made use of the VizieR catalogue access tool, CDS, Strasbourg, France (DOI : {\url{10.26093/cds/vizier}}). The original description of the VizieR service was published in 2000, A\&AS 143, 23. This research has made use of data and/or software provided by the High Energy Astrophysics Science Archive Research Center (HEASARC), which is a service of the Astrophysics Science Division at NASA/GSFC. This research has made use of the SDSS query service CASJOBS {\url{(http://skyserver.sdss.org/CasJobs/)}}, and the CasJobs is a service provided by the Johns Hopkins University.

LOFAR is the Low Frequency Array designed and constructed by ASTRON. It has observing, data processing, and data storage facilities in several countries, which are owned by various parties (each with their own funding sources), and which are collectively operated by the LOFAR ERIC under a joint scientific policy. The LOFAR resources have benefited from the following recent major funding sources: CNRS-INSU, Observatoire de Paris and Université d'Orléans, France; BMFTR, MKW-NRW, MPG, Germany; Science Foundation Ireland (SFI), Department of Business, Enterprise and Innovation (DBEI), Ireland; NWO, The Netherlands; The Science and Technology Facilities Council, UK; Ministry of Science and Higher Education, Poland; The Istituto Nazionale di Astrofisica (INAF), Italy. 

This research made use of the Dutch national e-infrastructure with support of the SURF Cooperative (e-infra 180169) and the LOFAR e-infra group. The Jülich LOFAR Long Term Archive and the German LOFAR network are both coordinated and operated by the Jülich Supercomputing Centre (JSC), and computing resources on the supercomputer JUWELS at JSC were provided by the Gauss Centre for Supercomputing e.V. (grant CHTB00) through the John von Neumann Institute for Computing (NIC).

This research made use of the University of Hertfordshire high-performance computing facility and the LOFAR-UK computing facility located at the University of Hertfordshire and supported by STFC [ST/P000096/1], and of the Italian LOFAR-IT computing infrastructure supported and operated by INAF, including the resources within the PLEIADI special "LOFAR" project by USC-C of INAF, and by the Physics Department of Turin university (under an agreement with Consorzio Interuniversitario per la Fisica Spaziale) at the C3S Supercomputing Centre, Italy.

This research is part of the project LOFAR Data Valorization (LDV) [project numbers 2020.031, 2022.033, and 2024.047] of the research programme Computing Time on National Computer Facilities using SPIDER that is (co-)funded by the Dutch Research Council (NWO), hosted by SURF through the call for proposals of Computing Time on National Computer Facilities.

\bibliography{winged2}{}
\bibliographystyle{aasjournal}

\appendix
\section{Sample Polar Maps and PA Measurements}
\label{appendix_PA_measurements}
Here we present a set of representative WRGs whose wing structures are less well defined and whose lobes and wings exhibit noticeable asymmetries. These examples illustrate cases where PA measurements carry higher uncertainty, although such complexities do not significantly affect the global statistical results.

\begin{figure*}[ht!]
\includegraphics[angle=0,width=3.80cm]{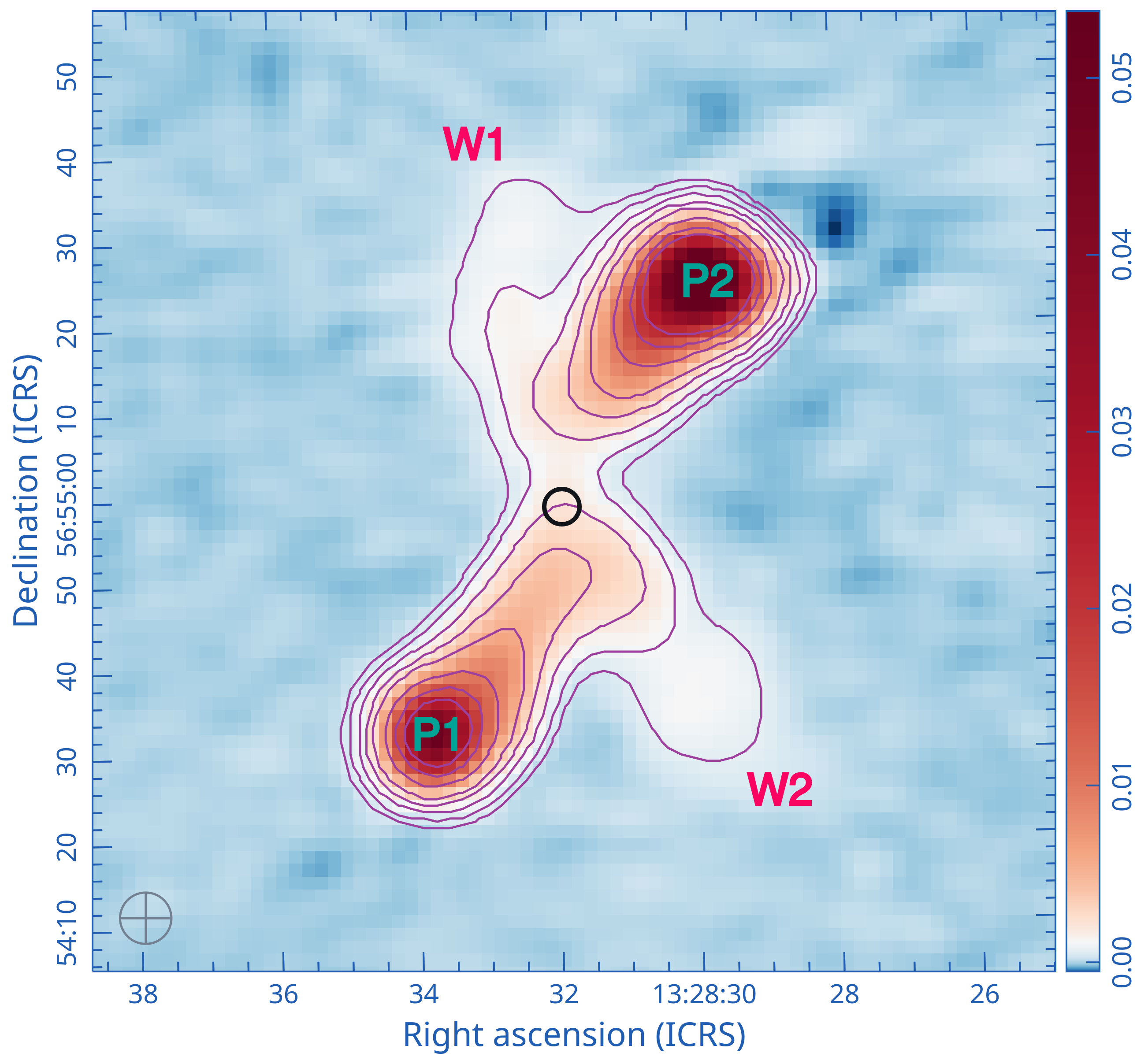}
\includegraphics[angle=0,width=13.70cm]{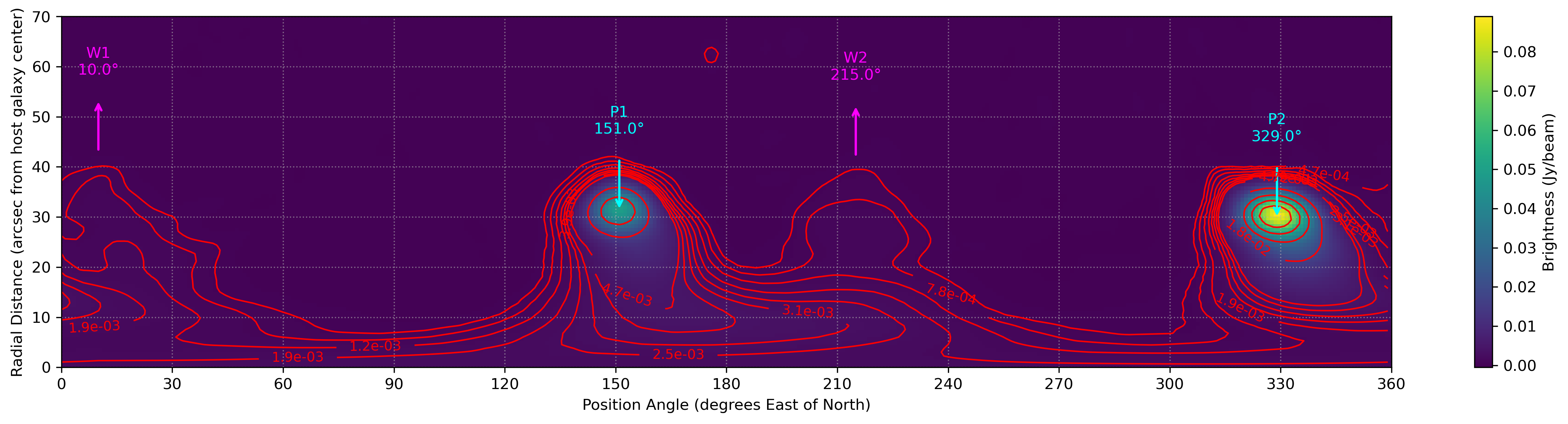}
\includegraphics[angle=0,width=3.80cm]{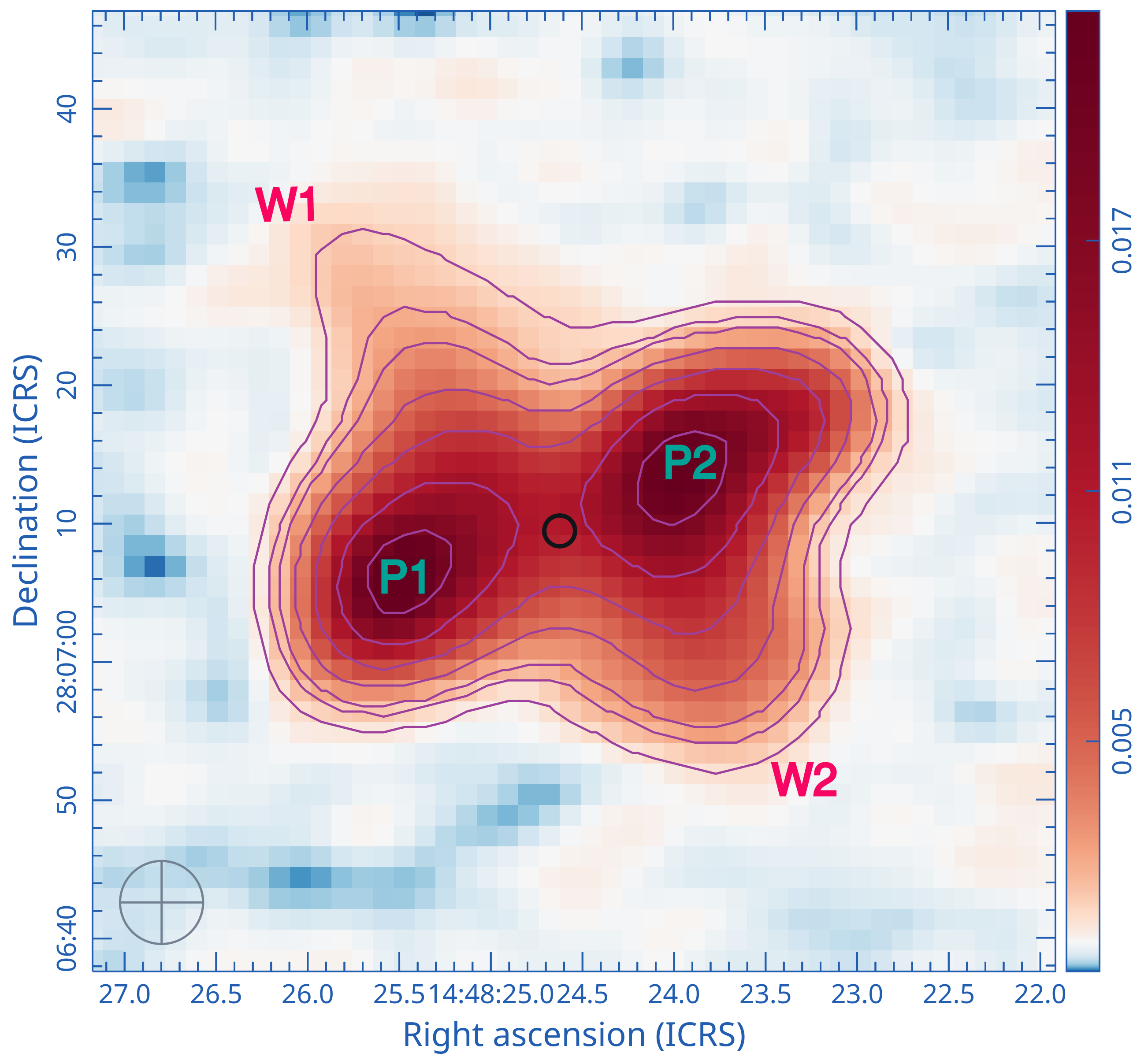}
\includegraphics[angle=0,width=13.70cm]{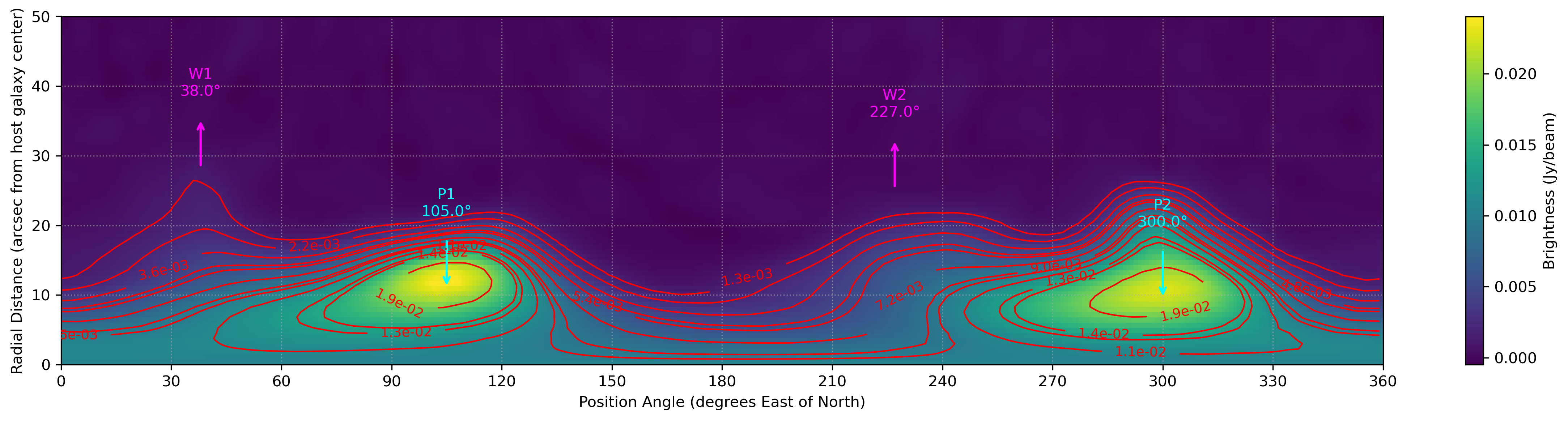}
\includegraphics[angle=0,width=3.80cm]{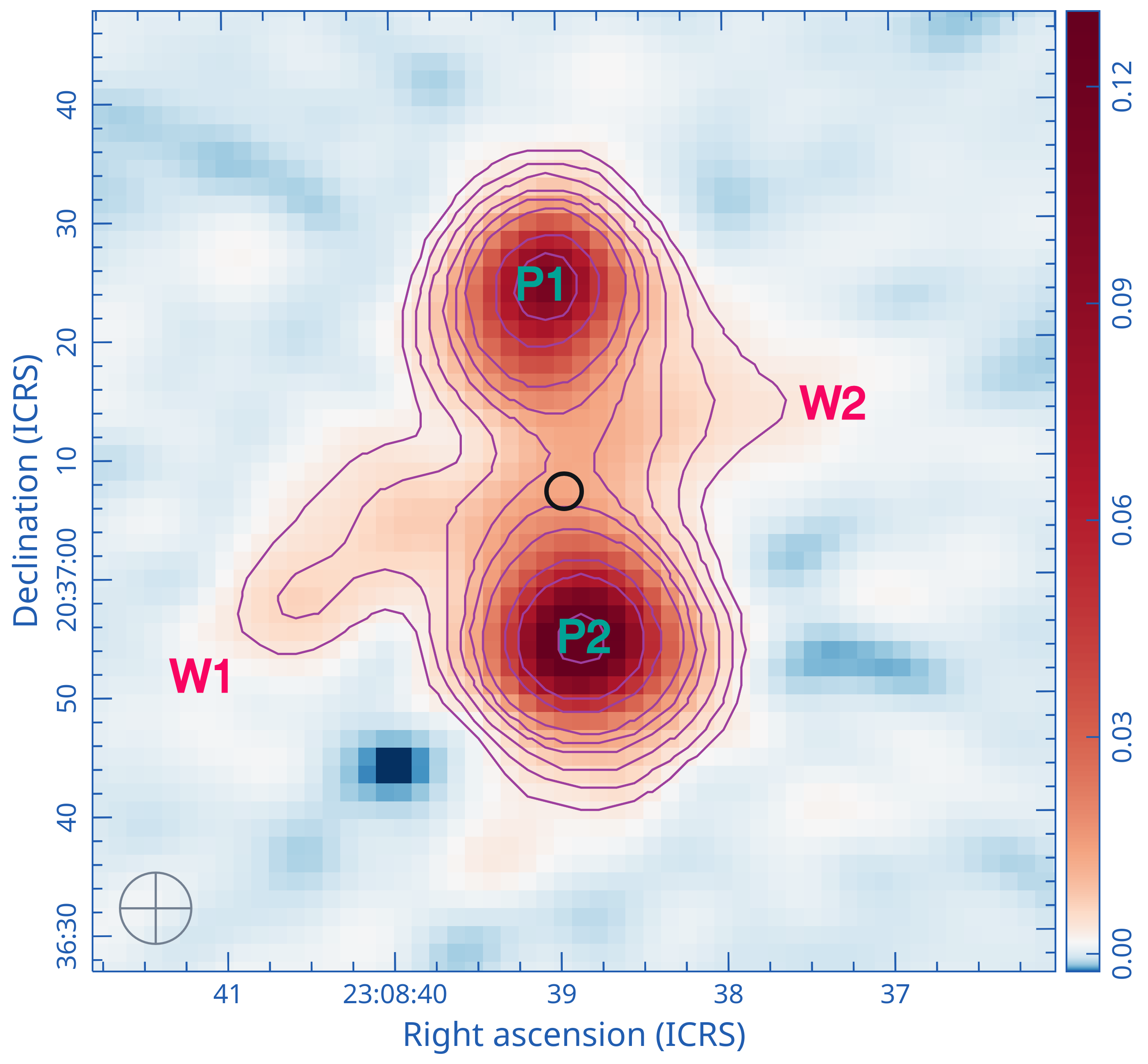}
\includegraphics[angle=0,width=13.50cm]{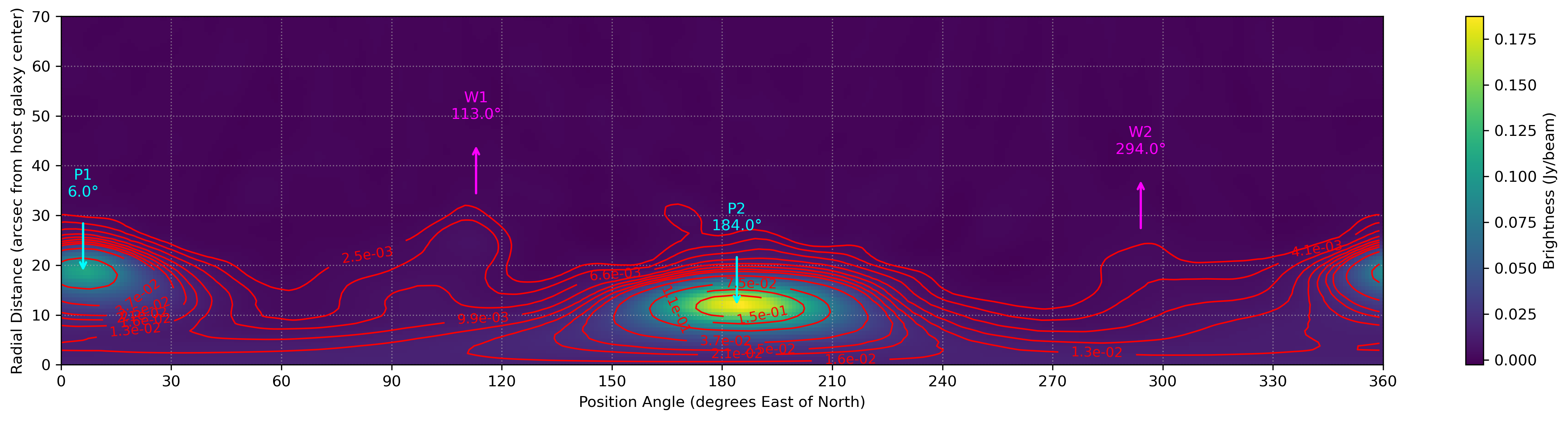}
\includegraphics[angle=0,width=3.80cm]{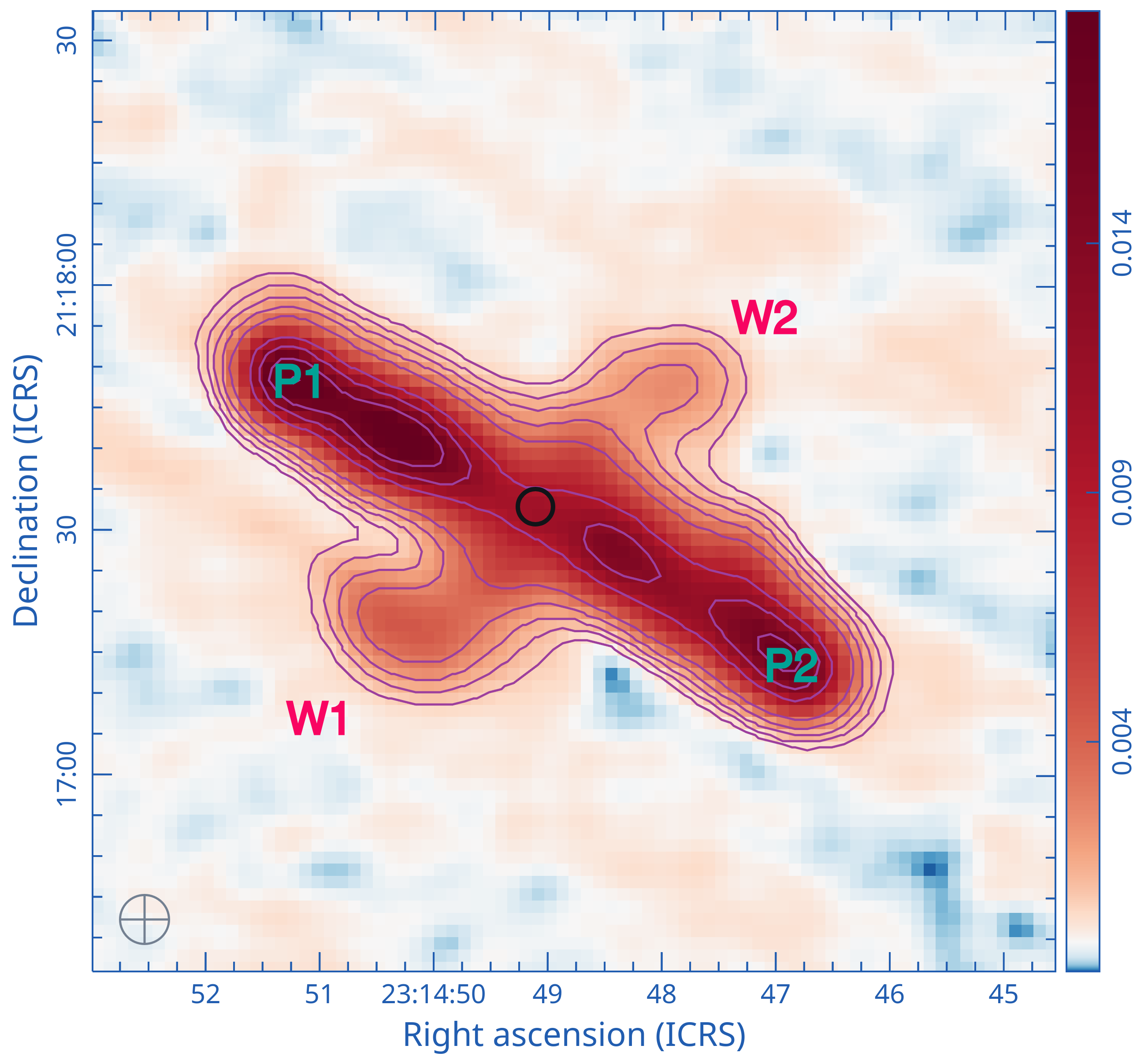}
\includegraphics[angle=0,width=13.70cm]{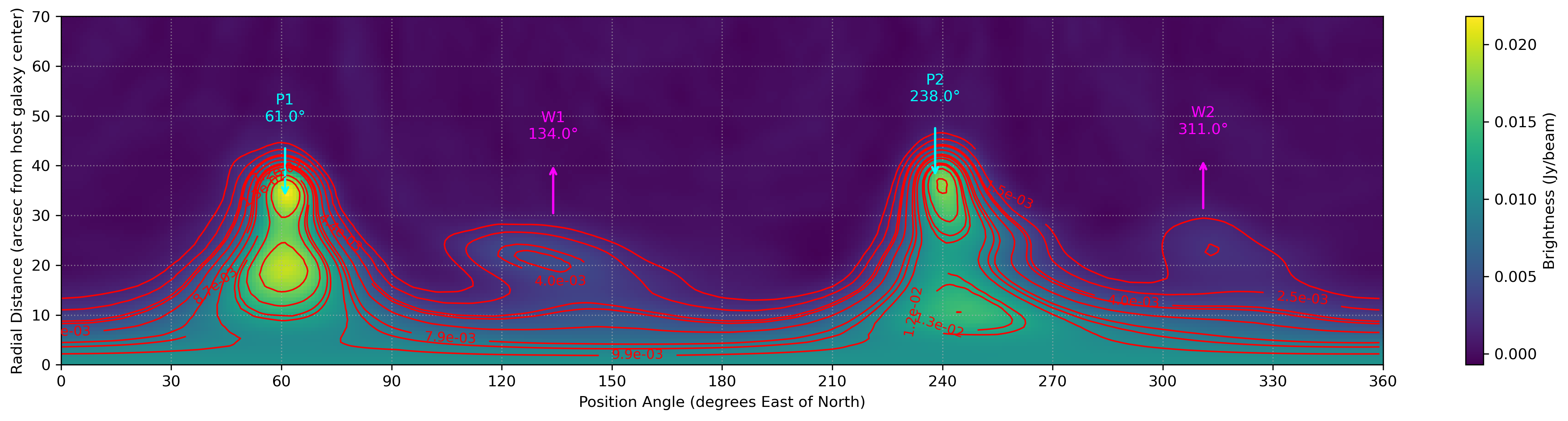}
\includegraphics[angle=0,width=4.40cm]{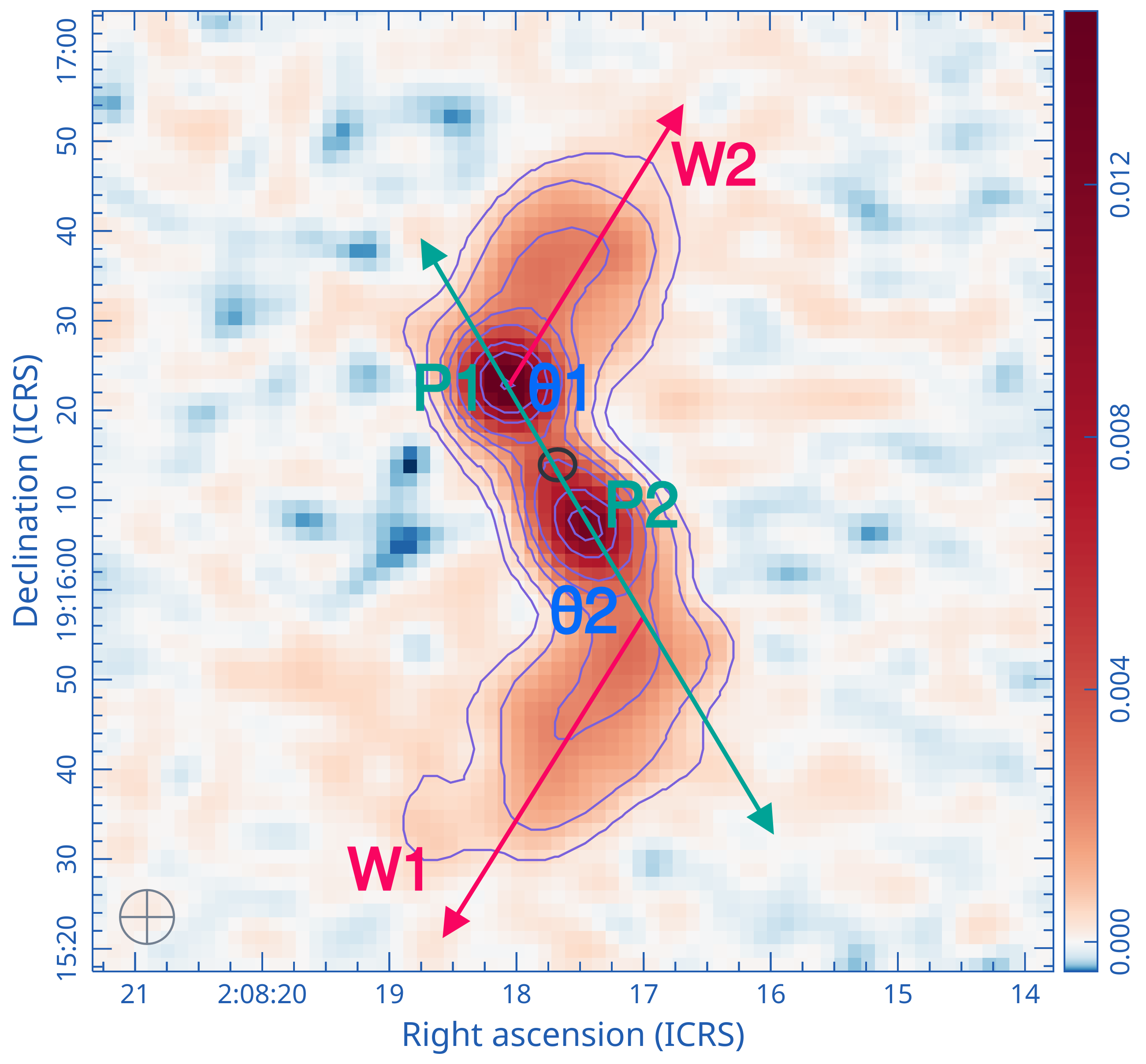}
\includegraphics[angle=0,width=4.40cm]{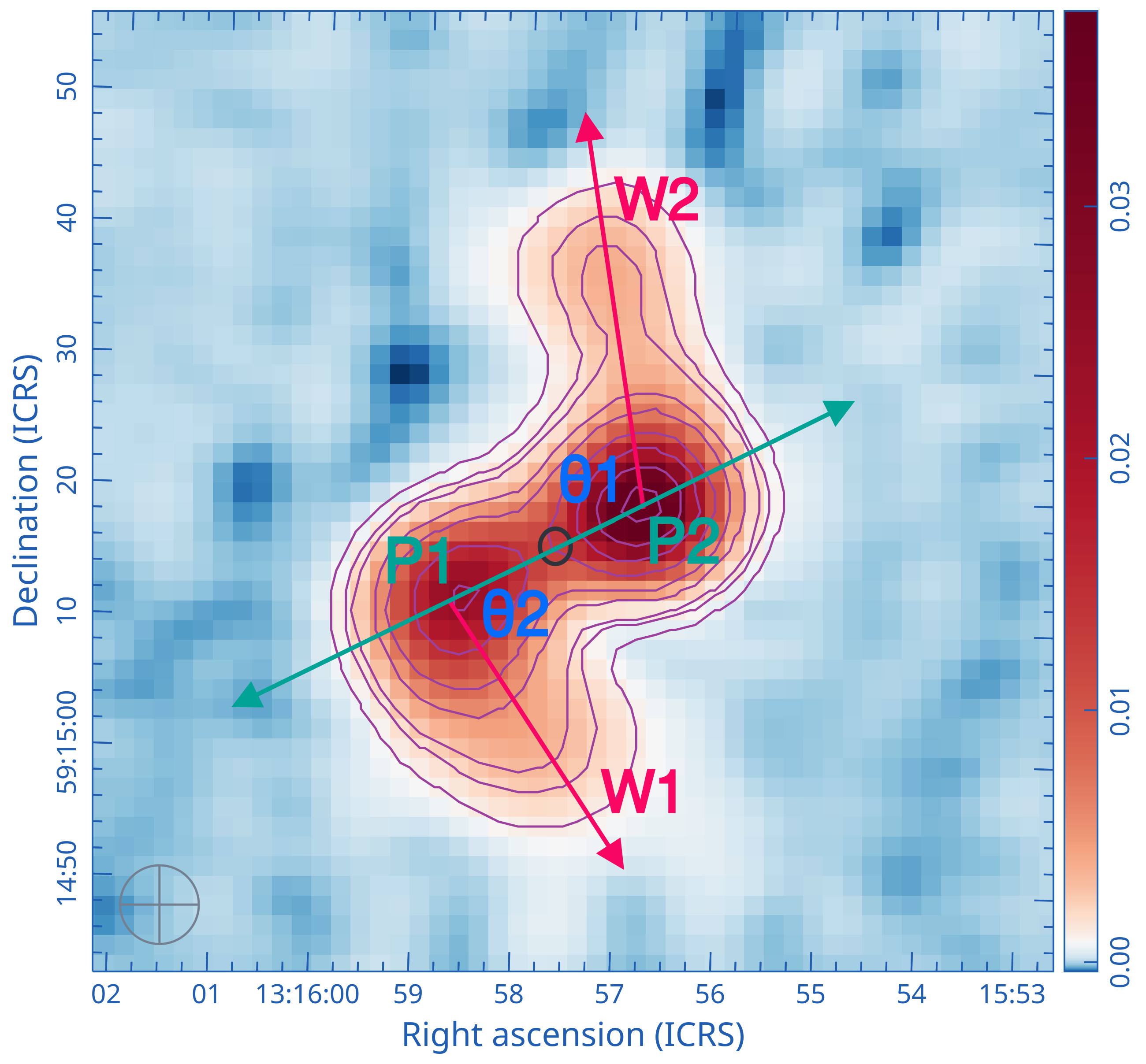}
\includegraphics[angle=0,width=4.40cm]{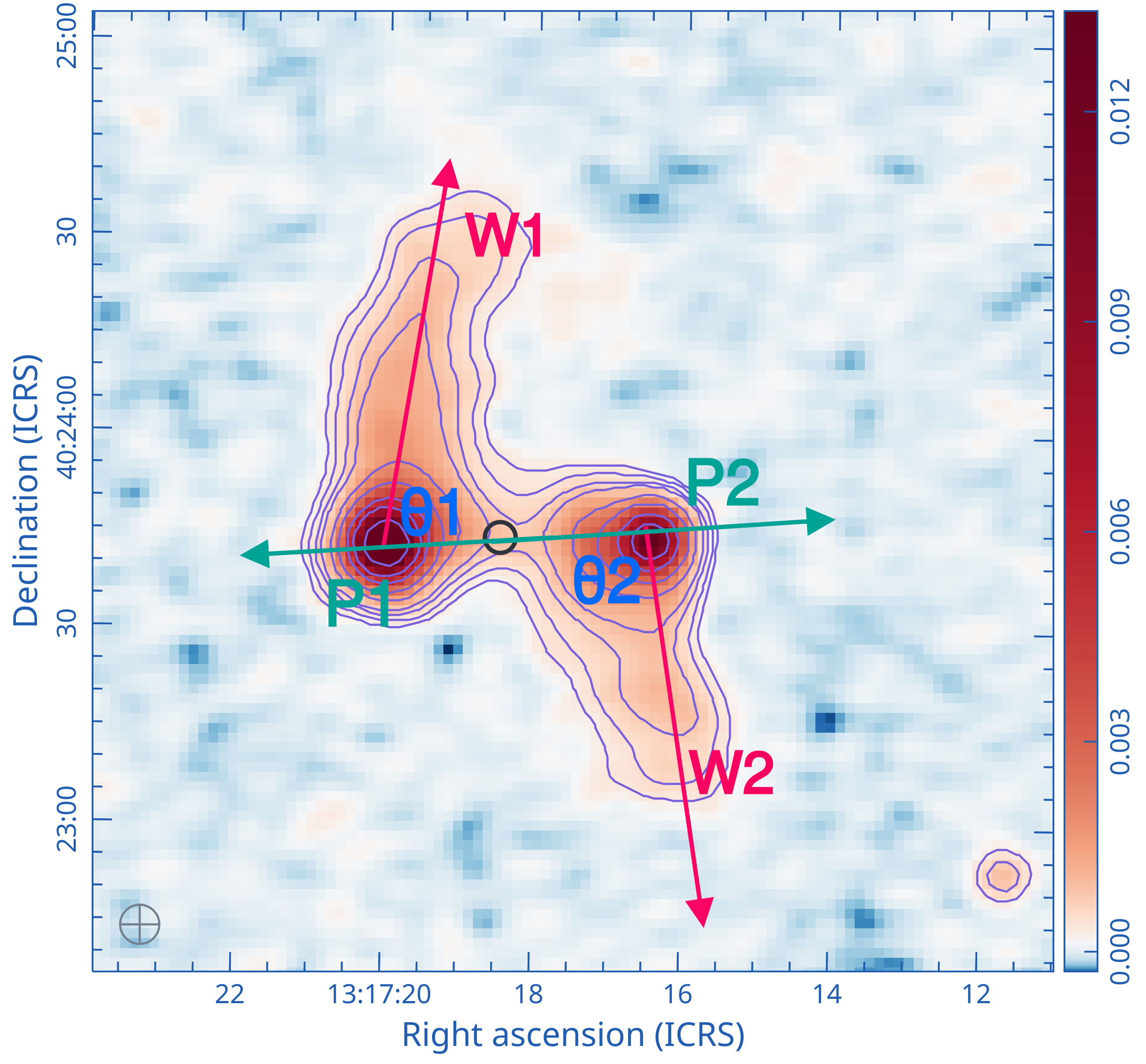}
\includegraphics[angle=0,width=4.40cm]{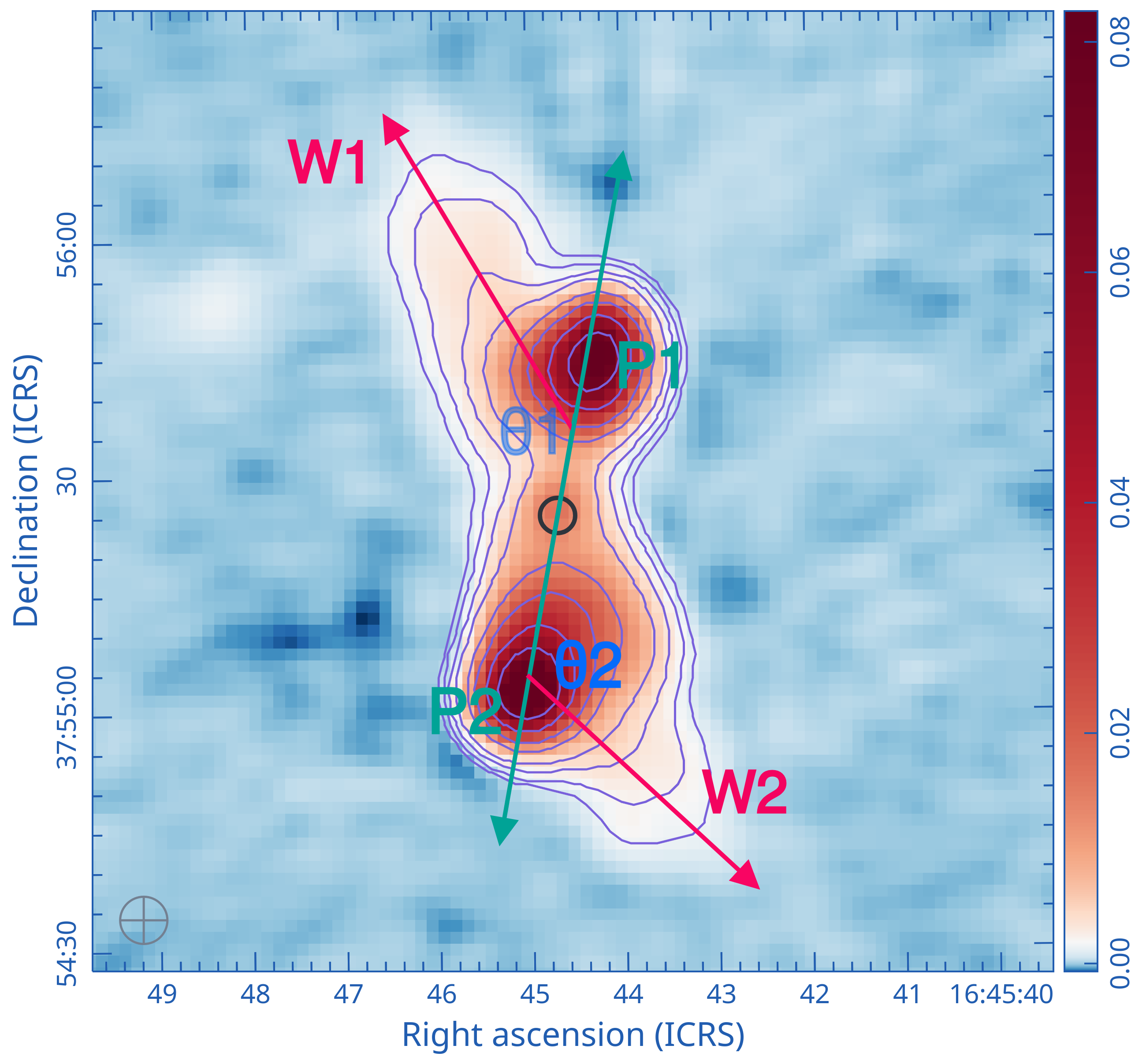}
\label{fig:radio_PA2}
\caption{Eight representative examples illustrating the polar-diagram transformation and position-angle (PA) measurements for XRGs, and the wing–primary lobe offset angle measurements for ZRGs. A set of four XRGs (J1328+5654, J1448+2807, J2308+2037, and J2314+2117) and ZRGs (J0208+1916, J1315+5915, J1317+4023, and J1645+3755) are shown. XRGs (first five sources): The left-hand panels show the LoTSS DR2 radio images, with the primary lobes (P1, P2) and wings (W1, W2) identified, and the host galaxy marked by a circle. The right-hand panels display the corresponding polar maps constructed by adopting the host galaxy as the origin. The measured positions and PAs are indicated by inward-pointing cyan arrows for the primary lobes and outward-pointing magenta arrows for the wings. ZRGs (last four sources): Schematic diagrams illustrating the wing–primary lobe offset angle measurements ($\rm \theta$1, $\rm \theta$2), defined between the primary jets (P1, P2) and the corresponding wings (W1, W2).}
\end{figure*}

\section{Cross-Validation of Position Angle Measurements with Literature}
\label{appendix_PA_verification}
To assess the reliability of our position angle (PA) measurements, we performed a direct cross-comparison with the sample of 22 sources analysed by \citet{2016A&A...587A..25G}. We first compared the measured wing PAs and found an excellent agreement between the two studies, with a Pearson correlation coefficient of 0.998 and a mean absolute difference of 3.27$^{\circ}$.

\begin{figure*}[ht!]
\centering
\includegraphics[angle=0,width=9.0cm]{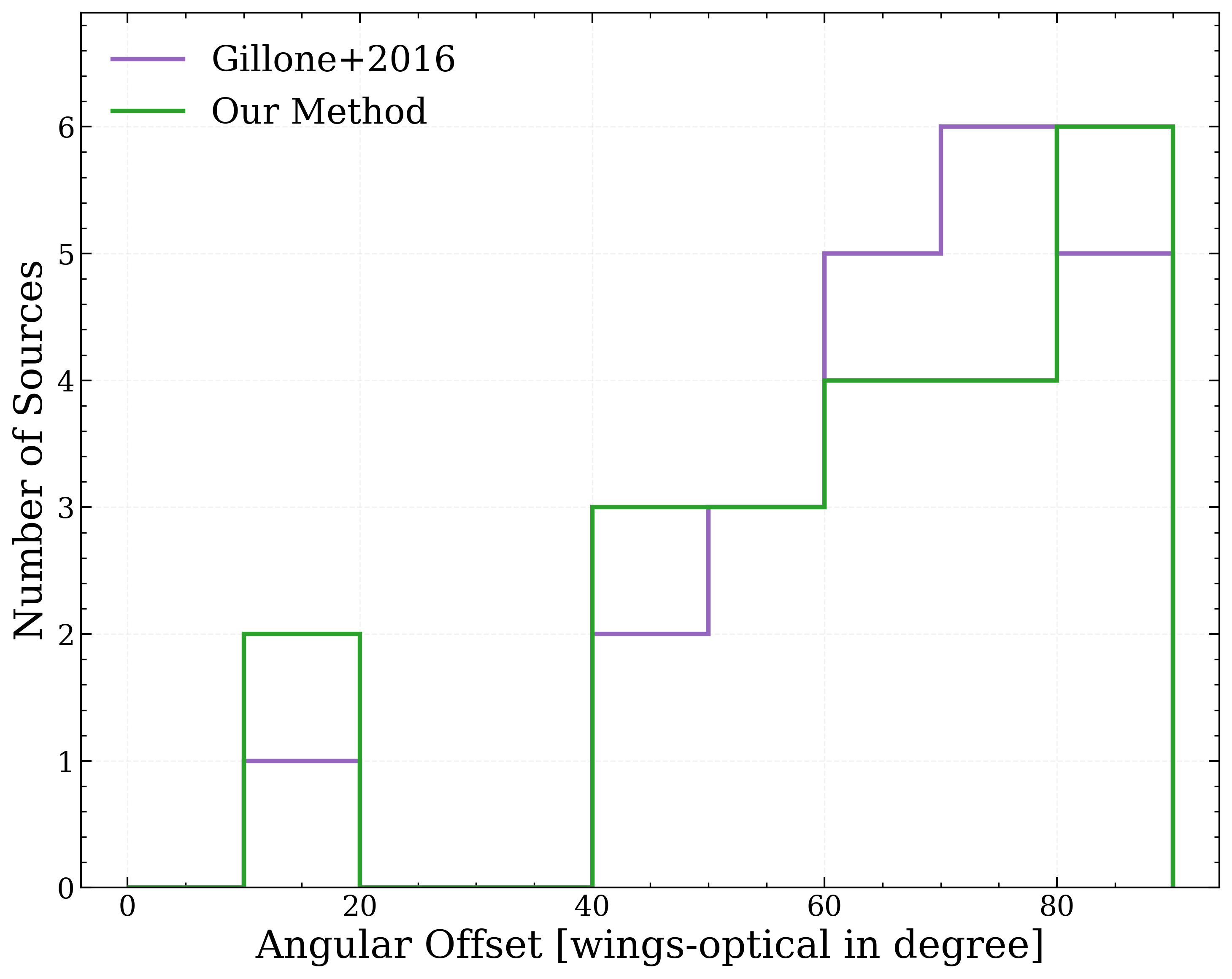}
\label{fig:compare_Gillone}
\caption{Comparison of offset PA measurements. The purple line corresponds to the \citet{2016A&A...587A..25G} measurement, when the green one follows our measurement method.}
\end{figure*}

We then compared the offsets between the wing PAs and the optical major-axis PAs. These offsets also show good overall consistency with the benchmark study, with a mean absolute difference of 10.45$^{\circ}$. In Figure~\ref{fig:compare_Gillone}, we reproduce the offset PA distributions reported by \citet{2016A&A...587A..25G} alongside those obtained using our methodology. While the general shapes of the distributions are similar, small discrepancies are apparent.

Given the very strong agreement in the radio PA measurements, we attribute these differences primarily to the determination of the optical position angles. Indeed, we find a mean absolute difference of 9.86$^{\circ}$ between the optical PAs derived in the two studies. This is expected, as \citet{2016A&A...587A..25G} measured optical PAs by fitting elliptical isophotes to optical images using the IRAF {\it ellipse} task, whereas we employed two-dimensional Sérsic profile fitting. The isophotal method traces local surface-brightness contours, while the Sérsic fitting provides a more stable estimate of the global orientation of the stellar distribution.

Overall, this comparison demonstrates that our radio PA measurements are robust and that the modest differences in PA offsets arise from methodological differences in the optical PA estimation rather than from inconsistencies in the radio analysis.

\section{X-ray Properties of WRGs with Archival Detections}
\label{appendix_xray_table}
The following table lists WRGs with possible X-ray counterparts identified within 30$^{\prime\prime}$ of the host position from the available major archival X-ray catalogs. The tabulated quantities are taken directly from the respective catalogs.

\begin{deluxetable*}{lccccc}
\tablecaption{Archival X-ray Detections of the WRGs}
\tablewidth{0pt}
\tablehead{
\colhead
{WRG} &X-ray Source Name&X-ray      & Flux                                  & Energy Band &Source \\ 
ID No.&                 &Catalog    &($\times$ E-14 erg cm$^{-2}$ s$^{-1}$) & (keV)       &Extent \\
}
\startdata
~~13 &~~2CXO J001800.3+300623   &~~~~~~~CSC & 2.28$^{+2.64}_{-1.93}$ & ~0.5--7.0  &  F   \\	
~~29 &~4XMM J003411.6+393620    &XMMSSC     &  9.22$\pm$2.15       &  0.2--12.0 &  0   \\
~~31 &~~~2RXS J003530.7+222913  &~~~~~2RXS  &  68.07$\pm$25.26     &  0.1--2.4  &  0   \\
104  &1eRASS J075441.5+283210   &~~eRASS1   & 12.59$\pm$4.31       & ~0.2--0.6  &  0   \\
108  &~~~2RXS J081137.3+483124  &~~~~~2RXS  &  31.58$\pm$8.42      &  0.1--2.4  &  0.5 \\
117  &~~~2RXS J082119.5+251903  &~~~~~2RXS  &  20.35$\pm$7.71      &  0.1--2.4  &  0.5 \\
121  &1eRASS J082807.0+393540   &~~eRASS1   & 22.79$\pm$5.49       & ~0.2--0.6  &  0   \\
173  &~4XMM J100015.4+504445    &XMMSSC     &  9.07$\pm$3.45       &  0.2--12.0 &  42.4\\
192  &~~2CXO J104320.5+585620   &~~~~~~~CSC & 1.51$^{+1.72}_{-1.29}$ & ~0.5--7.0  &  F   \\
212  &1eRASS J111333.2+354225   &~~eRASS1   & 7.55$\pm$2.98        & ~0.2--0.6  &  0   \\
239  &~4XMM J121812.4+504500    &XMMSSC     &  9.08$\pm$9.77       &  0.2--12.0 &  0   \\
246  &1eRASS J122409.7+283513   &~~eRASS1   & 18.24$\pm$4.20       & ~0.2--0.6  &  0   \\
277  &~4XMM J124622.4+263236    &XMMSSC     & 0.59$\pm$0.35        &  0.2--12.0 &  0   \\
309  &~~~2RXS J131132.2+394312  &~~~~~2RXS  &  17.54$\pm$6.31      &  0.1--2.4  &  0   \\
332  &~4XMM J132831.9+565500    &XMMSSC     &  28.90$\pm$4.32      &  0.2--12.0 &  0   \\
341  &~~~2RXS J133657.3+654116  &~~~~~2RXS  &  43.51$\pm$8.42      &  0.1--2.4  &  0   \\
342  &~~~2RXS J133745.4+415049  &~~~~~2RXS  &  16.84$\pm$6.31      &  0.1--2.4  &  0.2 \\
355  &~~~2RXS J134944.4+344134  &~~~~~2RXS  &  17.54$\pm$7.71      &  0.1--2.4  &  0   \\
365  &~4XMM J135614.4+383543    &XMMSSC     &  0.742$\pm$0.534     &  0.2--12.0 &  0   \\
376  &~~~2RXS J140656.2+582202  &~~~~~2RXS  &  60.35$\pm$9.82      &  0.1--2.4  &  0.1 \\
392  &~~2CXO J142434.5+453902   &~~~~~~~CSC &101.0$^{+106.0}_{-96.6}$& ~0.5--7.0  &  T   \\
396  &~~2CXO J142620.6+344016   &~~~~~~~CSC & 0.00$^{+0.191}_{-0.00}$& ~0.5--7.0  &  F   \\     
420  &~~~2RXS J145133.0+335737  &~~~~~2RXS  &  10.53$\pm$4.91      &  0.1--2.4  &  0   \\
424  &~~~2RXS J145546.4+361412  &~~~~~2RXS  &  29.47$\pm$7.01      &  0.1--2.4  &  0   \\
466  &~~~2RXS J154221.2+564805  &~~~~~2RXS  &  9.82$\pm$3.50       &  0.1--2.4  &  0   \\
523  &~~~2RXS J164545.0+375514  &~~~~~2RXS  &  18.95$\pm$7.01      &  0.1--2.4  &  0   \\
584  &~~~2RXS J221537.4+290241  &~~~~~2RXS  &  75.09$\pm$11.23     &  0.1--2.4  &  0.6 \\
591  &~~~2RXS J224508.4+325909  &~~~~~2RXS  &  23.86$\pm$7.01      &  0.1--2.4  &  0   \\
\enddata
\tablecomments{The above table has the following columns: Column (1): WRG id number (as in \hyperlink{cite.2025ApJS..278...34B}{Paper I}). Column (2): X-ray source name as listed in the original survey catalog. Column (3): X-ray catalog from which the source was identified (2RXS, CSC, XMMSSC, or eRASS1). Here, CSC v2.1.1 is used for CSC, and 4XMM DR14 is used for XMMSSC. Column (4): Observed X-ray flux. Fluxes are given in units of erg cm$^{-2}$ s$^{-1}$. For 2RXS sources, the count rates were converted to fluxes using \hyperlink{https://heasarc.gsfc.nasa.gov/cgi-bin/Tools/w3pimms/w3pimms.pl}{PIMMS} assuming a generic spectral model (power law) with photon index 1.7 and galactic hydrogen column density 0.1 cm$^{-2}$, in the 0.1 -- 2.4 keV band. The values are given with their errors, except for CSS sources, where the values are given with the lowest and highest limits. Column (5): Energy band in which the flux was measured. Column (6): Source extent information, as reported in the catalogs, where the zero corresponds to the point-like source and the non-zero values indicate extended emission. For CSC catalog sources, the flag `T' is marked for extended or inconsistent with point source (90\% confidence).}
\end{deluxetable*}

\end{document}